\documentclass{sig-alternate-10pt}
\usepackage{graphicx}
\usepackage[justification=centering]{caption}
\usepackage[format=hang]{subcaption}
\usepackage{url}
\usepackage{amsfonts}
\usepackage{times}
\usepackage{threeparttable}
\usepackage{amsmath, amssymb}
\usepackage{pslatex}
\usepackage{enumerate}
\usepackage{url}
\usepackage{arcs}
\usepackage{yhmath}
\usepackage{epstopdf}

\usepackage[toc,page]{appendix}
\usepackage{color} 

\usepackage{booktabs}

\usepackage[ruled,vlined]{algorithm2e}
\usepackage{multirow,array}
\usepackage[nocompress]{cite}
\SetAlFnt{\fontsize{8.75}{8.75}}
\usepackage{hyphenat}
\usepackage{enumitem}

\setlist[itemize]{leftmargin=*}
\setlist[enumerate]{leftmargin=*}

\usepackage[ngerman,british]{babel}
\useshorthands{"}
\makeatletter
\declare@shorthand{ngerman}{";}{\discretionary{-}{-}{-}\bbl@allowhyphens}
\makeatother
\addto\extrasbritish{\languageshorthands{ngerman}}

\newcommand{\qinsays}[2][]{}
\newcommand{\yysays}[1]{\hspace{0pt}}
\newcommand{\dbsays}[1]{}

\newcommand{\sysn}{Concury\xspace}
\newcommand*\xor{\mathbin{\oplus}}
\usepackage{authblk}
\begin{document}
	\newcommand{\mysubsubsection}[1] {
		\subsubsection{#1}
	}
	\renewcommand{\paragraph}[1]{
		
		\noindent \textbf{#1}
		
	}
	\newcommand{\superscript}[1]{\ensuremath{^{\textrm{#1}}}}
	\newcommand{\CONEXT}[1]{\textcolor{blue}{#1}}
	\def\sharedaffiliation{\end{tabular}\newline\begin{tabular}{c}}
	
	\newcommand{\widthinTriColumn}{0.9\linewidth}
	\newcommand{\indentinbox}{\noindent \hangafter=1 \setlength{\hangindent}{1em}}

	\date{}
	

\title{\LARGE \bf Concury: A Fast and Light-weighted Software Load Balancer}

\author[1]{Shouqian Shi}
\author[1]{Chen Qian}
\author[2]{Ye Yu}
\author[1]{Xin Li}
\author[3]{Ying Zhang}
\author[4]{Xiaozhou Li}
\affil[1]{University of California, Santa Cruz}
\affil[2]{Google}
\affil[3]{Facebook}
\affil[4]{Celer Network}

	\maketitle
	
	
	\subsection*{Abstract}
	A load balancer (LB) is a vital network function for cloud services to balance the load amongst resources. Stateful software LBs that run on commodity servers provides flexibility, cost-efficiency, and packet consistency. However current designs have two main limitations: 1) states are stored as digests which may cause packet inconsistency due to digest collisions; 2) the data plane needs to update for every new connection, and frequent updates hurt throughput and packet consistency.
	In this work, we present a new software stateful LB called Concury, which is the first solution to solve these problems. 
	The key innovation of Concury is an algorithmic approach to store and look up large network states with frequent connection arrivals, which is succinct in memory cost, consistent under network changes, and incurs infrequent data plane updates.
	The evaluation results show that the Concury algorithm provides 4x throughput  and consumes less memory compared to other LB algorithms, while providing weighted load balancing and false-hit freedom, for both real and synthetic data center traffic. We implement Concury as a prototype system deployed in CloudLab and show that the throughput of Concury on a single thread can reach 62.5\% of the maximum capacity of two 10GbE NICs and that on two threads can reach the maximum capacity.
%

\section{Introduction}
\label{sec:intro}

A load balancer (LB) is a fundamental network function of a data center that provides Internet services.
In traditional cloud environments, to accommodate high demand for popular service at scale, such as a search engine, email, photo sharing/storage, or message posting and interactions, a data center maintains multiple backend servers, each carrying a direct IP (DIP).
For a particular service,  clients send their requests to a publicly visible IP address, called the virtual IP (VIP). There could be many DIPs \emph{behind} one VIP. An LB uses different DIPs to replace the VIP on the service requests and balances the load across the servers, so that no server gets overloaded to disrupt the service. An LB usually operates on or above layer 4.

Conventional hardware-based LBs \cite{Netscaler,LBorg,A10} have limitations on scalability, availability, flexibility, and cost-efficiency \cite{Maglev}. Hence major web services such as Google \cite{Maglev}, Microsoft \cite{Ananta}, and Facebook \cite{FBSLB} has started to rely on software stateful LBs, which scale by using a distributed data plane that runs on commodity servers, providing high availability, flexibility, and cost-efficiency.
The key functions of a stateful LB include the following. 1) For a \emph{stateless packet}, which can be sent to an arbitrary DIP supporting its VIP, the LB algorithm should act as a \emph{weighted randomizer} based on the current capacities of the backend servers. 2) For a \emph{stateful packet}, which belongs to a connection including prior packets forwarded to a DIP, the LB forwards it to the particular DIP, rather than an arbitrary one, preserving \emph{packet consistency} (also called per-connection consistency (PCC) \cite{SilkRoad})

The major challenge of a stateful LB algorithm is to preserve packet consistency under network dynamics, including new connection arrivals and DIP changes due to server failures or updates. Most existing LB algorithms use hash tables to store connection states as the data plane solution \cite{Ananta,Maglev}.
These stateful LBs experience large memory cost of storing packet states or low capacity of packet processing. They require a large number of commodity servers to scale out, e.g., up to 3.5\% - 10\% of the data center size as reported by Microsoft \cite{duet} and Google \cite{Maglev}.
Hence some LBs use digests of connections rather than full connection state (such as 5-tuples) to reduce memory cost and improve throughput \cite{Maglev}. This design has two major weaknesses: 1) using digests may cause violation of packet consistency due to digest collisions; 2) massive numbers of new connections cause highly frequent data plane updates -- a modern cluster may easily experience thousands of new connections per second \cite{SilkRoad} -- which significantly hurt the packet processing throughput and possibly violate packet consistency.
Existing methods on fast and concurrent reads and writes to hash tables \cite{MemC3,Cuckoo-Li} cannot be easily applied to LB algorithms, because they only work with full state keys rather than digests. Recent work uses ASICs on programmable switches for fast table lookups \cite{SilkRoad}, but further increases the infrastructure cost.

We propose the first stateful LB algorithm that resolves the current limitations, called \sysn.\footnote{The name Concury is from Concordia, the Roman goddess of balance and harmony, and Mercury, the Roman god of messages/communication and travelers, known for his great speed.}
Its key innovation and contribution is a novel approach of \textbf{maintaining large-scale network states with a massive amount of newly arrival connections}, which is succinct in memory cost, consistent under network changes, and incurs extremely infrequent data plane updates. This approach could be possibly applied to many stateful network functions beyond LBs, such as NAT and EPC, but this paper only focuses on LBs.

We realize that the current limitations of software LBs are from the algorithmic designs for state maintenance and lookups: hash tables storing digests.
Hence \sysn resolves these challenges using two innovative ideas.
First, we use a data structure that represents all packet states
in a succinct manner (just two small arrays), by utilizing the theoretical foundation of MWHC minimal perfect hashing \cite{MonotoneHash,MWHC1996,BloomierFilter,Bloomier2nd2018,Concise_ICNP}.
A well-known variant of MWHC is the Bloomier filter (not Bloom filter) \cite{BloomierFilter,Bloomier2nd2018}.
\sysn is designed in such a way that it finds the specific destinations for stateful packets and \emph{simultaneously} acts as a weighted randomizer for stateless packets with small memory cost and packet consistency. 
Second,
we design the \sysn system including the coordination between the data and control planes such that \textit{\sysn does not need to update its lookup tables for every incoming connection}. Instead the  data plane of \sysn can be updated once every backend server change (DIP change), which happens much less frequently than new state arrivals: once every ten minute vs. thousands of times per second. State maintenance and updates in \sysn are much simpler than existing solutions,
which allow \sysn to maintain high lookup throughput and consistency.

In addition, \sysn can be  used for more complex Internet applications and the emerging edge cloud \cite{shi2016edge,Fog-QunLi15,FacebookEdge2017,GoogleEdge2017}. 1) It fits the condition of an Edge that typically  has constrained resource -- the LB in an Edge may only be hosted by few servers and could be co-located with other services on the servers \cite{shi2016edge,Fog-QunLi15,SDLB}.
2) Traditional cloud LBs consider a state for every TCP connection. However, in modern cloud or Edge, states may be for \emph{multi-connection} and at the device-level or process-level \cite{shi2016edge,Fog-QunLi15,SDLB}, i.e., the packets belonging to a same device should be sent to a same DIP. For example, a user device may keep offloading its video data to an edge server, let the server processes the data, and later request the analytical results from the server \cite{shi2016edge}. This whole process consists of multiple TCP flows and pseudo-flows (e.g., UDP), all of which should be sent to a consistent DIP. Unlike previous designs, the nature of \sysn can easily support multi-connection states. 3) Modern cloud and Edge servers might have heterogeneous capacity in computation, storage, and bandwidth \cite{shi2016edge,Fog-QunLi15}.
\sysn reacts quickly to the weight changes due to  failures or load dynamics of the servers.

\textbf{We make several key intellectual contributions}, in addition to using
MWHC hashing \cite{Concise_ICNP}, including:
\begin{enumerate}
\item The workflow of \sysn is designed to achieve memory-efficiency,
 high throughput,  load balancing, consistency, and false hit freedom. 
\item We invent a new data structure to maintain the dynamic set of states in the control plane and can instantly produce new lookup structures to update the data plane, under DIP pool changes.
\item We add the functions of weighted randomizer and maintaining multi-connection state to LBs.
\item We implement \sysn using DPDK \cite{DPDK} to demonstrate its high performance in a publicly-available network platform \cite{CloudLab}. We also build a P4 prototype to show its compatibility to programmable switches. We release the anonymous source code and our results can be  verified and re-produced \cite{ConcuryCode}.
\end{enumerate}

\textbf{In a nutshell, this work is indeed about improving  L4LB, which has been studied for the recent years. However throughout intelligent design and extensive optimization, we have developed the fastest, scalable, and  most accurate software LB solution in the state of the art: \sysn achieves the highest packet processing throughput and (one of the) lowest memory cost with 0 false hit. We consider \sysn is a major improvement rather than incremental work on existing solutions: It achieves the best of three worlds: performance, cost-efficiency, and consistency (correctness). It fits new applications and systems. It also includes an ideal solution of dynamic state maintenance  that is useful for other network functions.}

The balance of this paper is organized as follows. Section~\ref{sec:related} presents the related work and Section~\ref{sec:Othellobackground} introduces some background. We formally model the LB algorithm in Section~\ref{sec:problem}.  We present the detailed design of \sysn in Section~\ref{sec:design}.
The system implementation and evaluation results are shown in Section~\ref{sec:evaluation}. We conclude this work in Section~\ref{sec:conclusion}. This work does not raise any ethical issues.

\newcolumntype{C}[1]{>{\centering\let\newline\\\arraybackslash\hspace{0pt}}m{#1}}
\begin{table*}[ht]
 \centering \small
  \begin{tabular}{cC{1.8cm}C{1.5cm}C{1.5cm}C{1.5cm}C{1.5cm}C{1.3cm}C{1.3cm}C{1.5cm}}
     \toprule
  LB Algorithm     &Used in&Lookup speed&Memory cost&Weighted LB&False hits&Packet type&Extra hardware&Update interrupt\\
     \hline
 ECMP + hash table & Ananta\cite{Ananta} & low  & high  & unclear & No & any type & No&frequent\\
     \hline
Hash table w/ digest & Maglev\cite{Maglev}  & moderate & moderate  & Yes & exist & TCP only & No&frequent\\
     \hline
Multi HTs w/ digest  & SilkRoad\cite{SilkRoad}  & high  & low  & No & exist & TCP only & ASIC &frequent\\
     \hline
\textbf{\sysn} (this work) & - & high & low  & Yes  & No & any type & No&infrequent\\

\bottomrule
\end{tabular}
  \small\caption{Qualitative comparison among stateful LB algorithms.}
\label{tbl:comp}
\end{table*}

\section{Related Work}
\label{sec:related}
An LB is an important component of a data center network, which distributes incoming traffic to different backend servers or other network functions \cite{Ananta,duet,openflowbalancing,Maglev,FBSLB}. Traditional hardware load balancers are expensive and not flexible.
Hence, many large cloud services
choose to use software load balancers \cite{Ananta,duet,Maglev,FBSLB,Nginx}. In addition, LBs are also important for edge data centers \cite{Fog-CCR,shi2016edge,SDLB,Araujo2018}, which allow heterogeneous devices on the path to the remote cloud to offer storage and computing resources.

\textbf{Stateful load balancers.} Ananta \cite{Ananta} is a software stateful LB in a three-level architecture.
However, each Ananta instance provides very slow packet processing speed as shown in \cite{duet}.  Duet \cite{duet} makes use of forwarding and ECMP tables on commodity switches to store VIP-DIP mappings. 
Maglev \cite{Maglev} is Google's distributed software load balancer running on commodity servers. 
The core algorithm of Maglev is to use a hash table to store connections as digests for load balancing and a new consistent hashing algorithm for resilience to DIP pool changes.
SilkRoad \cite{SilkRoad} implements LB functions on state-of-the-art programmable switching ASICs, which requires more than 50MB SRAM. It supports high-volume traffic with low latency and preserves PCC. Deploying SilkRoad introduces extra hardware cost --each SilkRoad switch costs 10K USD and multiple switches are needed for every cluster. In addition, both Maglev and SilkRoad may include false hits during connection lookups, because they use digests rather than the complete state information.
False hits cause two main problems.
1) A packet will be forwarded to a DIP that does not provide the correct service of its VIP and then fails.
2) Multiple states may share a same digest in the table.
It is difficult to decide when to delete a digest. Deleting the digest of a finished state might terminate an active state, if their digests collide.
Hence the table may keep exploding  or some active states will be terminated. The typical data structure that can be used to maintain states in the above methods is Cuckoo Hashing \cite{cuckoo}. Bonomi \emph{et al.} proposed to use ACSMs to maintain dynamic network states \cite{ACSM}, but this method cannot be used for LBs.
We qualitatively compare \sysn with existing stateful LBs in Table~\ref{tbl:comp}.

\textbf{Stateless load balancers.} Beamer~\cite{Olteanu2018} and Faild \cite{Araujo2018} are recently proposed stateless LBs. Their forwarding logics do not store connection state but using a simple mapping algorithm (static or consistent hashing).
The end servers need to examine \emph{every} packet header to ensure that the packet is consistent with the state on this server. If not, the server performs overlay re-routing to the correct DIP. This method requires a modification on the network stack of every server to add extra network processing. The computation and memory overheads are thus transferred to the server side and on a per-packet basis. Overlay re-routing might not be a significant problem when states are short-term. However for multi-connection states that are long-term, stateless LBs may cause re-routing of most stateful packets, because after a duration the mapping algorithm would become very different.
Stateless LBs are more resilient to the SYN-flood attack, but it still cannot mitigate such attack.
\textbf{We do not intend to decide a clear victory between stateful and stateless LBs.}
The purpose of this work is to focus on improving the stateful LB design and leave the choice between stateful and stateless LBs to network operators.
\section{Background: Bloomier and Othello}
\label{sec:Othellobackground}

We propose to use the data structure and algorithms of MWHC minimal perfect hashing \cite{MonotoneHash,MWHC1996,BloomierFilter,Bloomier2nd2018,Concise_ICNP} for the \sysn LB.
One well-known perfect hashing based data structure  is the Bloomier filters \cite{BloomierFilter,Bloomier2nd2018}.
The recently proposed  Othello Hashing \cite{Concise_ICNP, Concise_ToN} makes use of Bloomier filters for the forwarding tables in the programmable networks, including a variant of Bloomier filters as its data plane,  a construction program in its control plane,  as the interaction protocols of the two planes.
It was not designed for LBs. Though Othello itself is not a merit of this work, we briefly describe it to help understanding how \sysn works.

Othello is used as a mapping for a set of key-value pairs.
Let $S$ be the set of keys and $n=|S|$.
The lookup of each key returns an $l$-bit value corresponding to the key.   

\begin{figure}[t]
    \centering
    \includegraphics[width=0.9\linewidth]{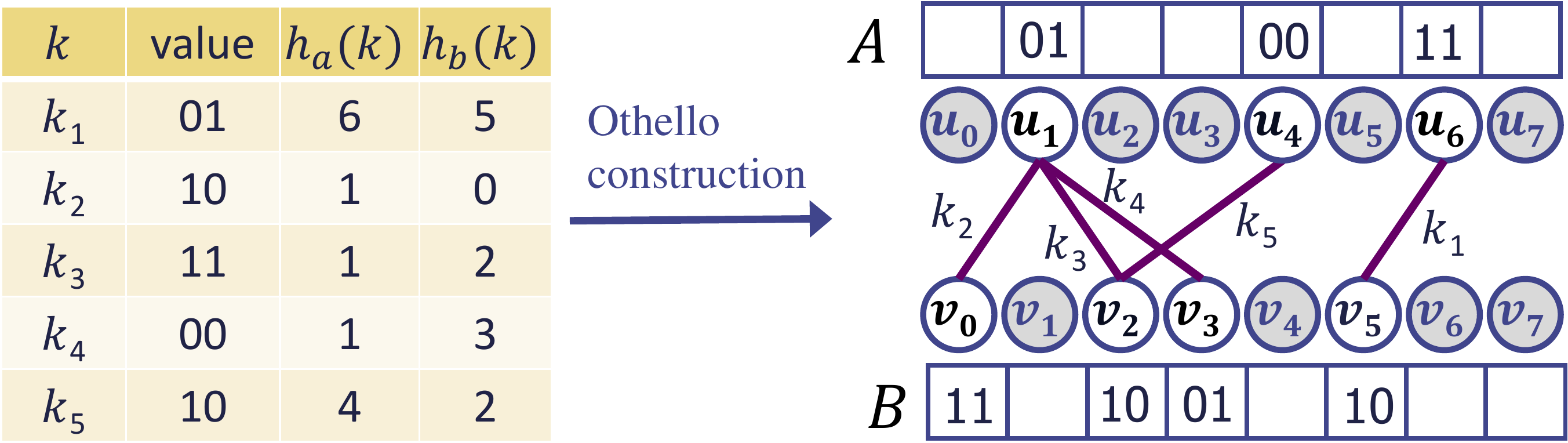}
    \caption{Construction of  Othello}
    \label{fig:OthelloConstL}
\end{figure}

\textbf{Othello construction in the control plane.}
We use an example in Fig.~\ref{fig:OthelloConstL} to show Othello Hashing of a set of five key-value pairs. Each of the keys $k_1$ to $k_5$  has a corresponding $l$-bit value. Two arrays $A$ and $B$ are built
with $m_a$ and $m_b$ elements respectively, $m_a, m_b >n$. Each element of an array is an $l$-bit value. In this example $m=m_a=m_b=8$. For every value $i$ in $A$ we place a vertex $u_i$ and for every value $j$ in $B$ we place a vertex $v_j$.
Othello uses two hash functions $h_a$ and $h_b$ and computes the integer hash values in $[0,m-1]$ for all keys.
Then, for each key, we place an edge between the two vertices that correspond to its hash values. For example, $h_a(k_1)=6$ and $h_b(k_1)=5$, so an edge is placed to connect $u_6$ and $v_5$.
For a key $k$ and its corresponding value $v$, the requirement of Othello is that the two connected elements $A[h_a(k)]\xor B[h_b(k)]=v$, where $\xor$ is the bit-wise \textit{exclusive or} (XOR).
For key $k_1$ in this example, $u_6 \xor v_5 = 10$. Vertexes colored grey represents ``not care'' elements.
Note after placing the edges for all keys, the bipartite graph, called graph $G$, needs to be \emph{acyclic}. It is proved that if $G$ is acyclic, it is trivial to find a valid element assignment such that the values of all keys are satisfied \cite{Concise_ICNP}. If a cycle is found, Othello needs to find another pair of hash functions to re-build $G$. It is proved that during the construction of $n$ keys, the expected total number of re-hashing is $<1.51$ when $n\leq0.75m$ \cite{Concise_ICNP}.
Hence the expected time cost to construct $G$ of $n$ keys is $O(n)$, and the expected time to add, delete, or change a key is O(1).

\begin{figure}[t]
    \centering
    \includegraphics[width=0.9\linewidth]{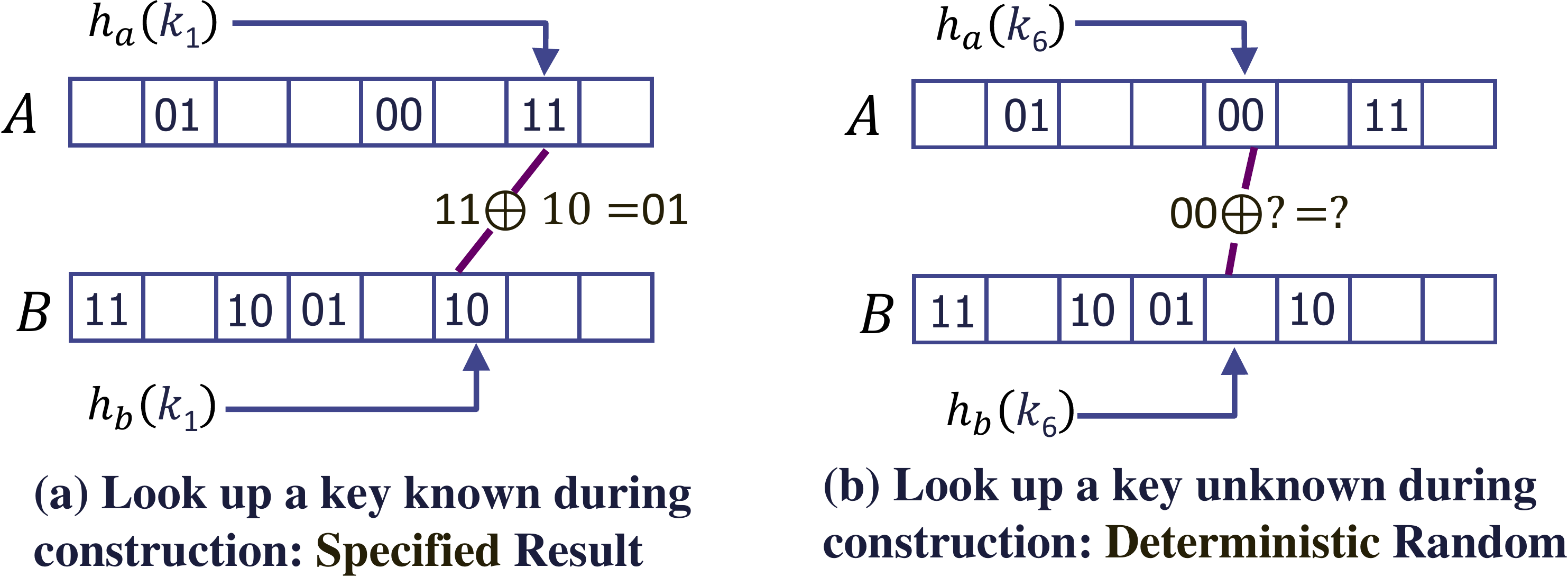}
    \caption{Lookups of Othello}
    \label{fig:OthelloLookup}
\end{figure}

\textbf{Othello lookups in the data plane.}
The  \emph{lookup structure} is simply the two bitmaps $A$ and $B$, as shown in Fig.~\ref{fig:OthelloLookup} (a).
To look up the value of $k_1$, we only need to compute $h_a$ and $h_b$, which are mapped to position 6 of $A$ and position 5 of $B$ (starting from 0). Then we compute the bit-wise XOR of the two bits and get the value 01. Hence the lookup result is $\tau(k)=a[h_a(k)]\oplus b[h_b(k)]$.

Othello lookups are memory-efficient and fast. 1) The data plane only needs to maintain the two arrays. The keys themselves are not stored in the arrays. Hence the space cost is small (2$m/n$ per key). 2) Each lookup costs just two memory access operations to read one element from each of $A$ and $B$. It fits the programmable network architecture: the data plane only needs to store the lookup structure, two arrays; the control plane stores the key-value pairs and the acyclic bipartite graph $G$. When there is any change, the control plane updates the two arrays and let the data plane to accept the new ones.

When Othello performs a lookup of a key that does not exist during construction, it returns an arbitrary value.
For example in Fig.~\ref{fig:OthelloLookup}(b), $k_6 \not\in S$ and its result may be an arbitrary value.
We will utilize this property to construct a \textbf{weighted randomizer}.

It should be noted that since Othello updates may require re-hashing, which, although happens in low probability ($O(1/n)$), still takes $O(n)$ which may introduce a notable latency to the control plane response time. Hence we propose an advanced data structure called OthelloMap that always maintain an up-to-date Othello structure in the control plane to limit the response time to ms level, as explained in \S~\ref{sec:OthelloMap}.

\section{System Models and Objectives}
\label{sec:problem}


A service provided by a cloud/edge data center is identified by a publicly visible IP address, called virtual IP (VIP). The clients send their service requests to the VIP.
An LB  balances the load across the cloud/edge servers, so that no server gets overloaded and disrupts the service. Each backend server is identified by a direct IP (DIP). Hence, the core function of an LB is to map the VIP on a packet header to a DIP, based on the header information of the packet (e.g., its 4-tuple or other state identifiers). 
Each VIP is associated with its \emph{DIP pool}, which includes the DIPs of the servers that provide the service identified by the VIP. The DIP pool of a VIP may vary depending on the service size and the environment (cloud or edge).
If a server maintains the state of packet, the packet must be sent to the DIP of the server. A state could be an ongoing connection or multi-connection.

Achieving all requirements of an LB stated in \S~\ref{sec:intro} is challenging. Simple stateless algorithms (such as static/consistent hashing) provides no guarantee of consistency. It is because the distribution algorithm needs to change when there is a DIP pool and weight change, and then stateful packets may be mapped to another server. Example of consistency violation by static hashing is shown in the Appendix A.

\begin{figure}[t]
    \centering
    \includegraphics[width=0.9\linewidth]{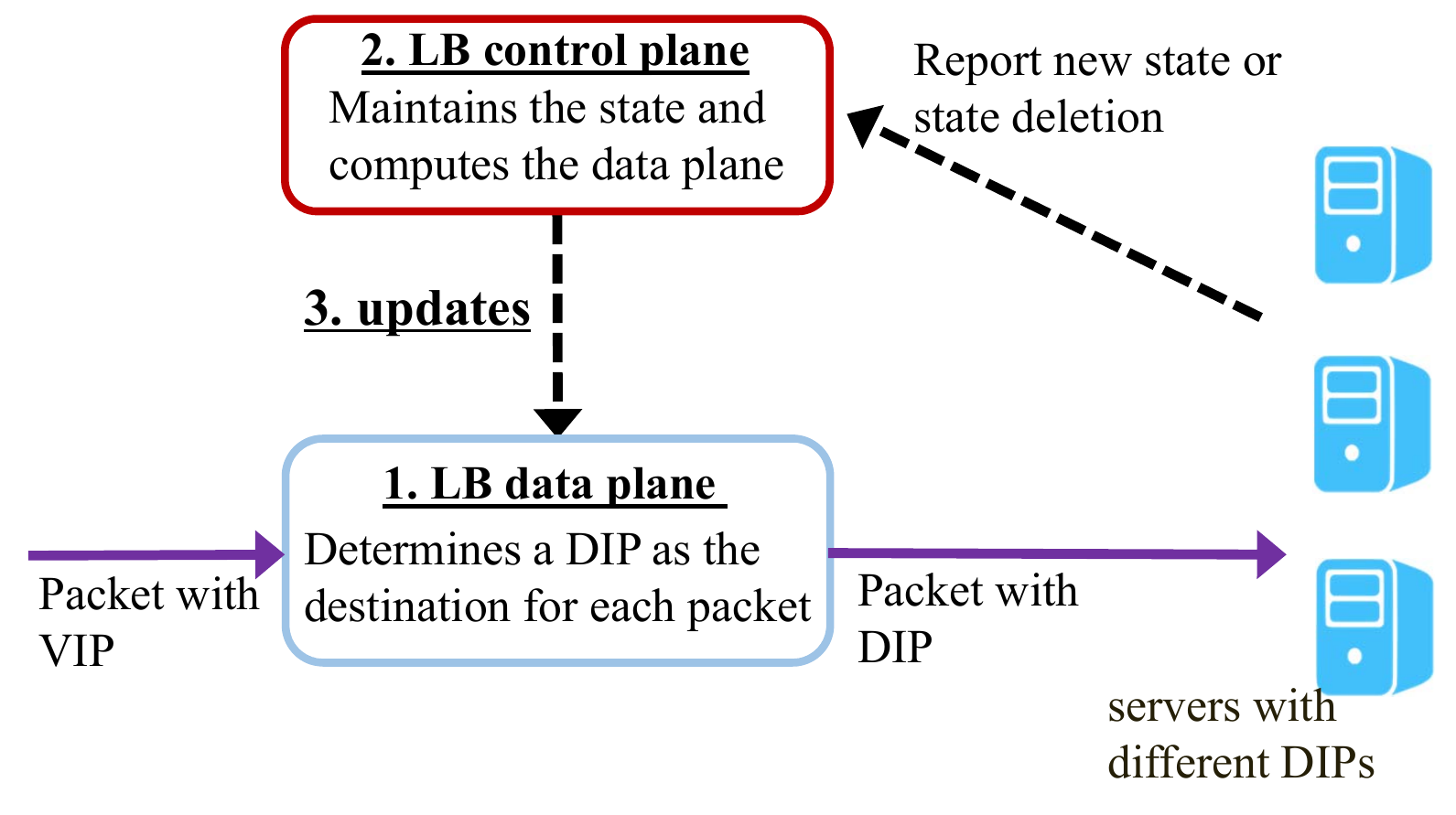}
    \caption{General model of a stateful LB}
    \label{fig:HashTable}
\end{figure}

Recent stateful LB designs \cite{Maglev,SilkRoad} need to store connection state and ensure that all packets matching a state are consistently mapped to a same DIP. We summarize a general model of stateful LBs. We analyze the components of this model and point out the algorithm design objectives.

\textbf{1. LB data plane (LB-DP).} The LB-DP processes packets and find a DIP for each packet carrying a VIP. The DIP should be selected from the DIP pool behind the VIP, representing the set of servers providing the service of this VIP. The core algorithm should provide two functions: i) find the corresponding server (DIP) for each stateful packet, and ii) assign an available server (DIP) based on given weights for each stateless packet. The design objectives of the LB-DP is to achieve \textbf{high packet processing throughput} and \textbf{efficiency of memory cost}, because high-speed memory is a precious resource on both commodity servers (cache) and hardware switches (ASICs). In addition, the LB-DP should \textbf{balance the stateless packets based on the weights} reflecting the current capacity of each server, which may be heterogeneous and dynamic. For example, if a server is serving many large-size connections, it has to receive fewer new states than others in the near future. So we identify the `weight' as an important input to the LB and we expect that an LB acts as a weighted randomizer for new states.

\textbf{2. LB control plane (LB-CP).} The LB-CP receives the state changes from the servers, including new state establishments and state removals. Many existing designs uses a TCP SYN packet as the indicator of a new state and allow LB-DP to notify the
LB-CP directly \cite{Maglev,SilkRoad}. However, it does not work for UDP or multi-connection states. The design objectives of the LB-CP is to efficiently \textbf{maintain all state} of the incoming packets and quickly construct the new LB-DP to reflect \textbf{packet consistency} once an LB-DP update is needed. Ensuring that all packets of a connection are delivered to the same server is critical for LBs, because recovering a broken connection usually takes a long time and significantly hurts the user experience.
Packets from a same device may also be sent to a same server because there may not be a unified data management layer in the edge and modern cloud. Achieving the device-level consistency could avoid overlay re-routing for many emerging applications such as media processing offloading.

\textbf{3. Update.} The LB-CP will notify LB-DP to make necessary changes under certain network dynamics, such as DIP pool and weight changes. The design objective of the update process is to \textbf{reduce the frequency of updating} because it will interrupt packet processing on the LB-DP.

Our design, implementation, analysis, and evaluation of \sysn will mainly focus on the above three components to achieve the stated objectives.

\section{Design of \sysn}
\label{sec:design}

\subsection{System overview}

\textbf{Notations.}
Let $M$ be the number of VIPs in the network. Each VIP $v_i$ is assigned an index $i$ and its DIP pool contains $t_i$ DIPs. The number of states of VIP $v_i$ is $n_i$.

\sysn follows the DP model introduced in \S~\ref{sec:problem}, including both data plane and control plane. The input of the \sysn data plane (\sysn-DP) is a packet whose destination address is a VIP and the output is the same packet whose destination has been replaced by a DIP. 
At each backend server (identified by a DIP), there is a host agent program that records the current states at this server, which has been already used in a prior work \cite{duet}. The host agent will report the \sysn control plane (\sysn-CP) about new and terminated states. \sysn-CP will update \sysn-DP only when the DIP pool of a VIP changes, e.g., server failure or addition. The update only applies to a small part of \sysn-DP.
The design objectives have been discussed in \S~\ref{sec:problem}.

\textbf{Challenges of designing \sysn.} One key innovation of \sysn is to abandon the conventional ``lookup-then-distribute'' workflow of prior LB designs and adopt a new approach that achieves `lookup' and `distribute' simultaneously. However, Othello Hashing was not originally designed for LBs. The challenges of applying Othello includes: 1)
how to adjust Othello for both active state lookups and weighted randomizer; 2) how to design the data plane to minimize memory cost and maximize throughput; 3) how to resolve the false hits problem without modifying the server network stack; and 4) how to relax the requirement of updating for every new state in the data plane.

\subsection{\sysn data plane}
\label{sec:dataplane}

\sysn uses Othello as both a lookup structure to represent the state-to-DIP mapping and a weighted randomizer. As introduced in \S~\ref{sec:Othellobackground}, an Othello lookup structure is built based on a set $S$ of keys. To apply Othello for \sysn, each key is the identifier of a state, i.e., 4-tuple. The value corresponding to a key is a DIP code ($Dcode$) which will be eventually converted to a DIP -- the address of a backend server that holds the state. Note the Othello lookup structure provides the state-to-DIP mapping but does not actually store the keys. Hence the memory cost is significantly reduced.

There are two possible approaches to construct the state-to-DIP lookup structure of \sysn. First, each VIP has an individual Othello lookup structure, which stores only the state-to-DIP mapping of this particular VIP. This requires $M$ Othellos in total.
We use this approach rather than a single and unified Othello due to the following reasons.

\begin{enumerate}
\item 
Under a change of a VIP's DIP pool, it is only necessary to update the Othello of \emph{this} VIP. The others are kept still.
\item Separate the lookups of different VIPs can ensure that packets are not forwarded to DIP in other VIP's pool.
\item Experimental results show that separate lookup structures provide faster lookup speed than a unified one.
\end{enumerate}
Note maintaining per-VIP structures can be also used by other stateful LBs such as Maglev \cite{Maglev}
to avoid the cross-VIP problem. However it still cannot resolve the digest-deletion problem stated in \S~\ref{sec:related}. \sysn is unique because it can deal with both types of problems.

\begin{figure}[t]
    \centering
    \includegraphics[width=0.95\linewidth]{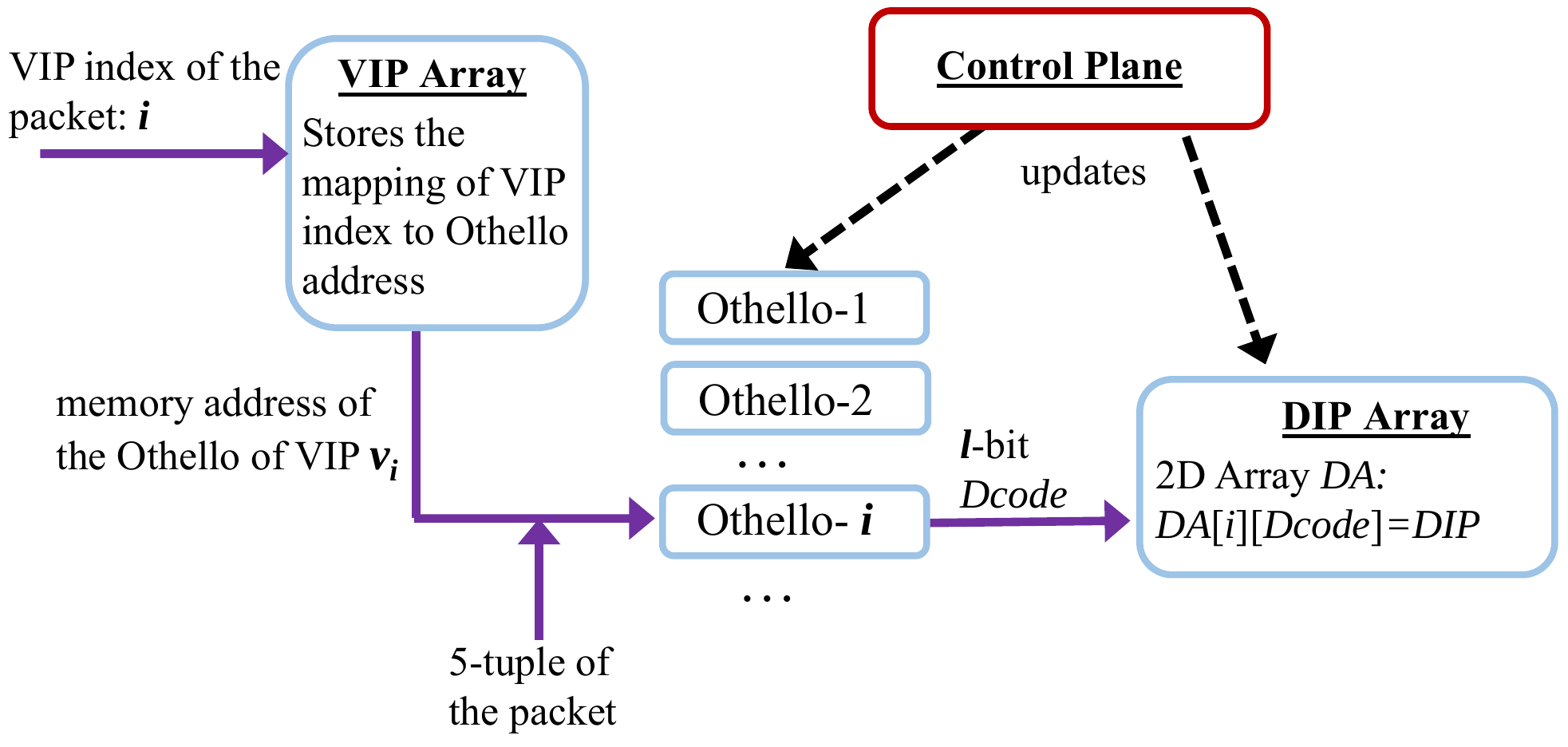}
    \caption{Workflow of \sysn data plane}
    \label{fig:ConcuryDP}
\end{figure}

The workflow of \sysn data plane is shown in Fig.~\ref{fig:ConcuryDP}, which includes three main steps.
We show the pseudocode in \emph{Appendix B}.
A unique property of \sysn is that the data plane lookup operation is \textit{simple and fast}, including just four read operations and the hash computation.

\textbf{Step 1.}
When \sysn receives a packet, it first gets the VIP index $i$ using the VIP $v_i$ in the packet header, by either a table lookup or calculation. Since VIPs are determined by the edge/cloud operator, one can simply assign all VIPs with a same prefix, e.g., 24-bit prefix. Then the last 8 bits of a VIP can be used as the VIP index, supporting 1K VIPs.
\sysn maintains a \emph{VIP array} that stores the memory addresses of different Othello lookup structures, using a static array whose index is the VIP index. The result of Step 1 is the memory address of the Othello of VIP $v_i$. The array is small and static.

\textbf{Step 2.} Using the memory address from Step 1, \sysn finds the Othello lookup structure for VIP $v_i$, denoted as Othello-$i$. Othello-$i$ only includes the two arrays $A$ and $B$ to support the calculation of the lookup result $\tau(k)=A[h_a(t)]\oplus B[h_b(t)]$, where $t$ is the 5-tuple. No 5-tuple information is stored.
The result is an $l$-bit value called DIP code, denoted as $Dcode$. Each DIP code will be mapped to an actual DIP in Step 3, and it is a many-to-one mapping. Two different DIP codes may be mapped to a same DIP.

\textbf{Step 3.} This step finds the actual DIP using the $l$-bit $Dcode$. \sysn maintains a 2D array called DIP array, denoted by $DA$. The element $DA[i][Dcode]$ is the DIP of the $Dcode$ for VIP $v_i$. This 2D array is independent of the number of current states and does not cost large memory. Assume there are 512 VIPs and $l=12$. The memory cost is about 2MB. Note  $DA[i][Dcode]$ for any $l$-bit value of $Dcode$ is a valid DIP of the VIP $v_i$. To further reduce the memory cost, $DA[i][Dcode]$ can be a DIP index will can be transferred to an actual DIP with one more static table lookup.

\textbf{Data plane complexity analysis and comparison.} Detailed analysis and comparison are presented in \emph{Appendix C}. Here we present the results.

\textbf{1) Time cost.} \sysn-DP is very simple and fast. Each lookup is in $O(1)$, including \textit{at most} 6 read operations from static arrays, 2 hash computations (32 bits for each), and an XOR computation. This cost is smaller than Cuckoo$+$digest, a commonly used LB table design \cite{Maglev,SilkRoad}, which needs more read operations and hash computations for both stateful and stateless packets.

\textbf{2) Space cost.} Let $n$ be the number of total states, $l_d$ be the length of Dcode, and $l_v$ be the length of the DIP index in the DIP table. The total memory cost of \sysn-DP is $2.33l_dn+64m+2^{l_d}l_vm+48*2^{l_v}$ bits, which is much smaller than that of Cuckoo$+$digest in practical setups.
In addition, compared to digest-based solutions, \sysn achieves space efficiency without causing false hits.

We will show the experimental results in \S~\ref{sec:evaluation}.

\begin{figure*}[t!]
\centering
\begin{tabular}{p{116pt}p{116pt}p{116pt}p{116pt}}
     \centering\includegraphics[width=\linewidth]{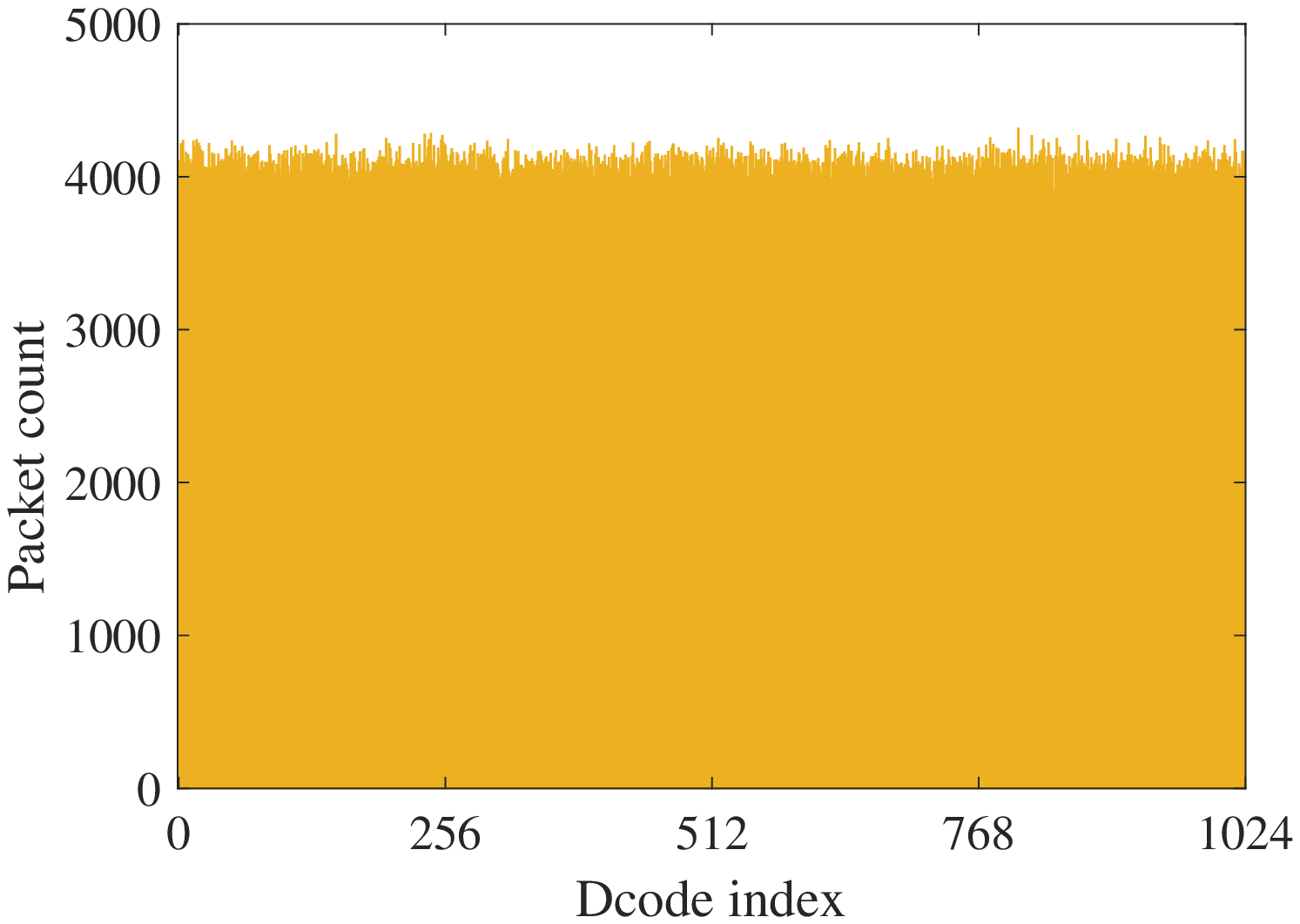}
     \caption{\small Stateless packet distribution by Dcode}
    \label{fig:1024-bucket}
&\centering\includegraphics[width=\widthinTriColumn]{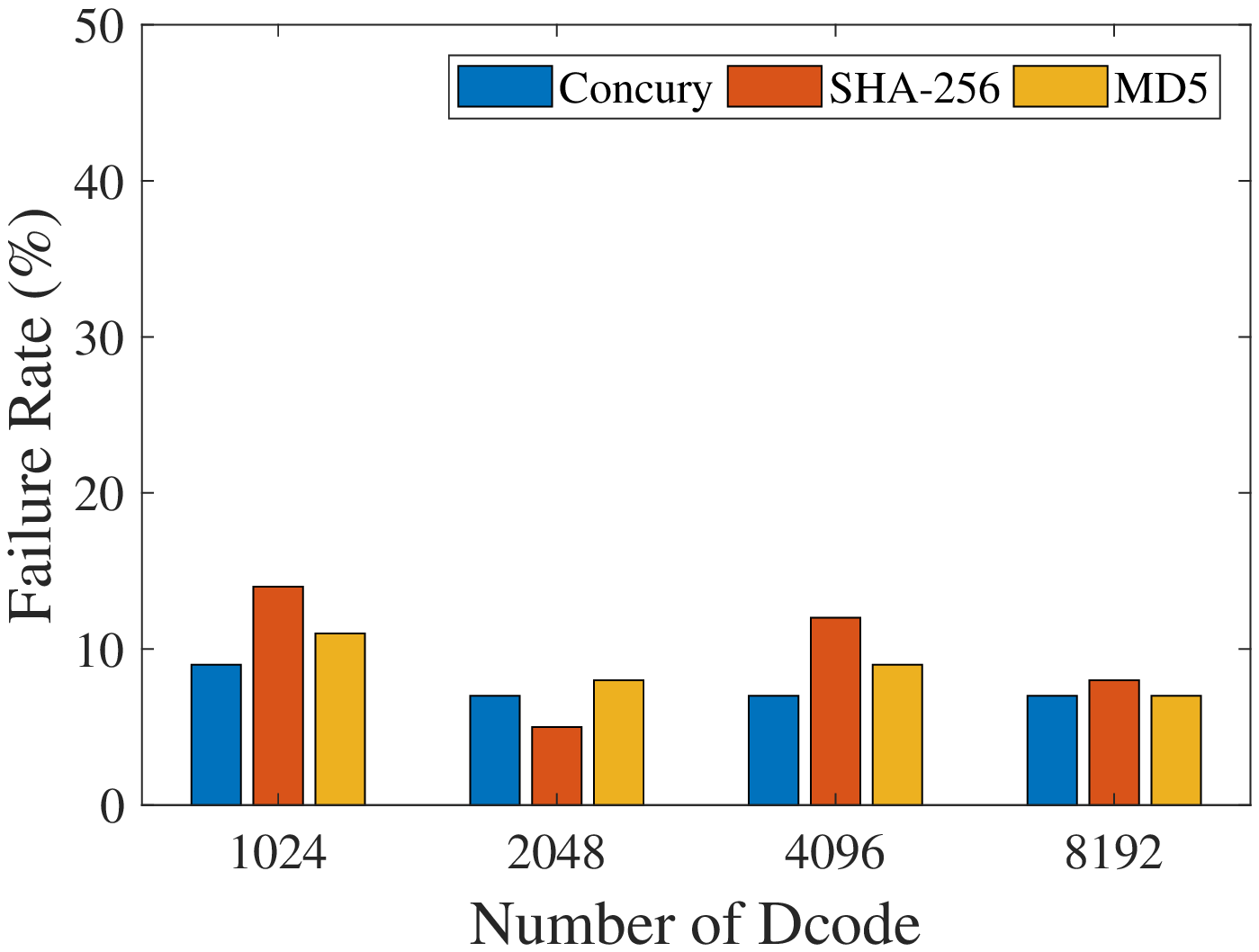}
    \caption{\small Chi-squared test}
    \label{fig:chi2-fail-bar}
&
    \centering\includegraphics[width=\widthinTriColumn]{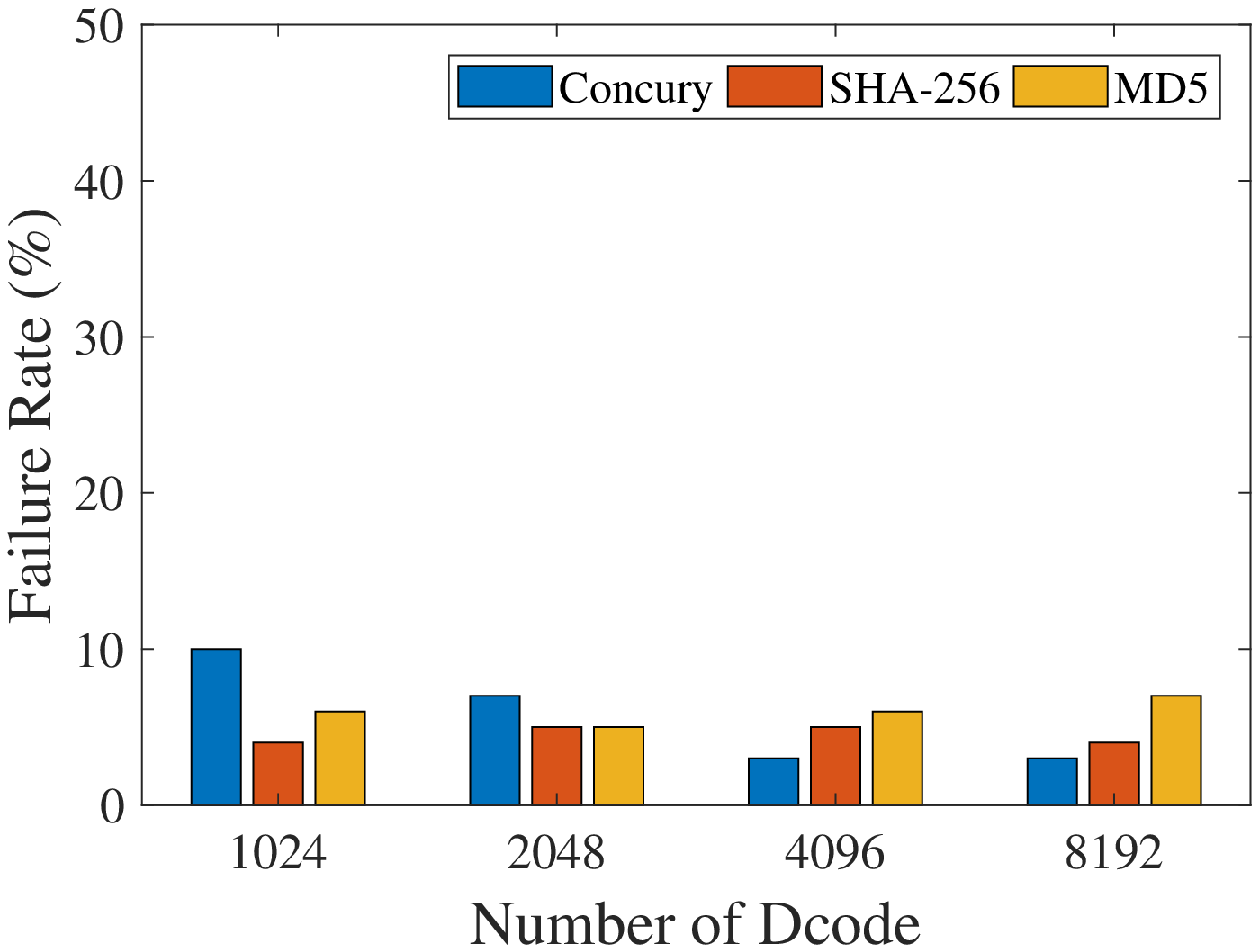}
    \caption{\small  Kolmogorov-Smirnov test}
    \label{fig:ks-fail-bar}
&
    \centering
    \includegraphics[width=0.95\linewidth]{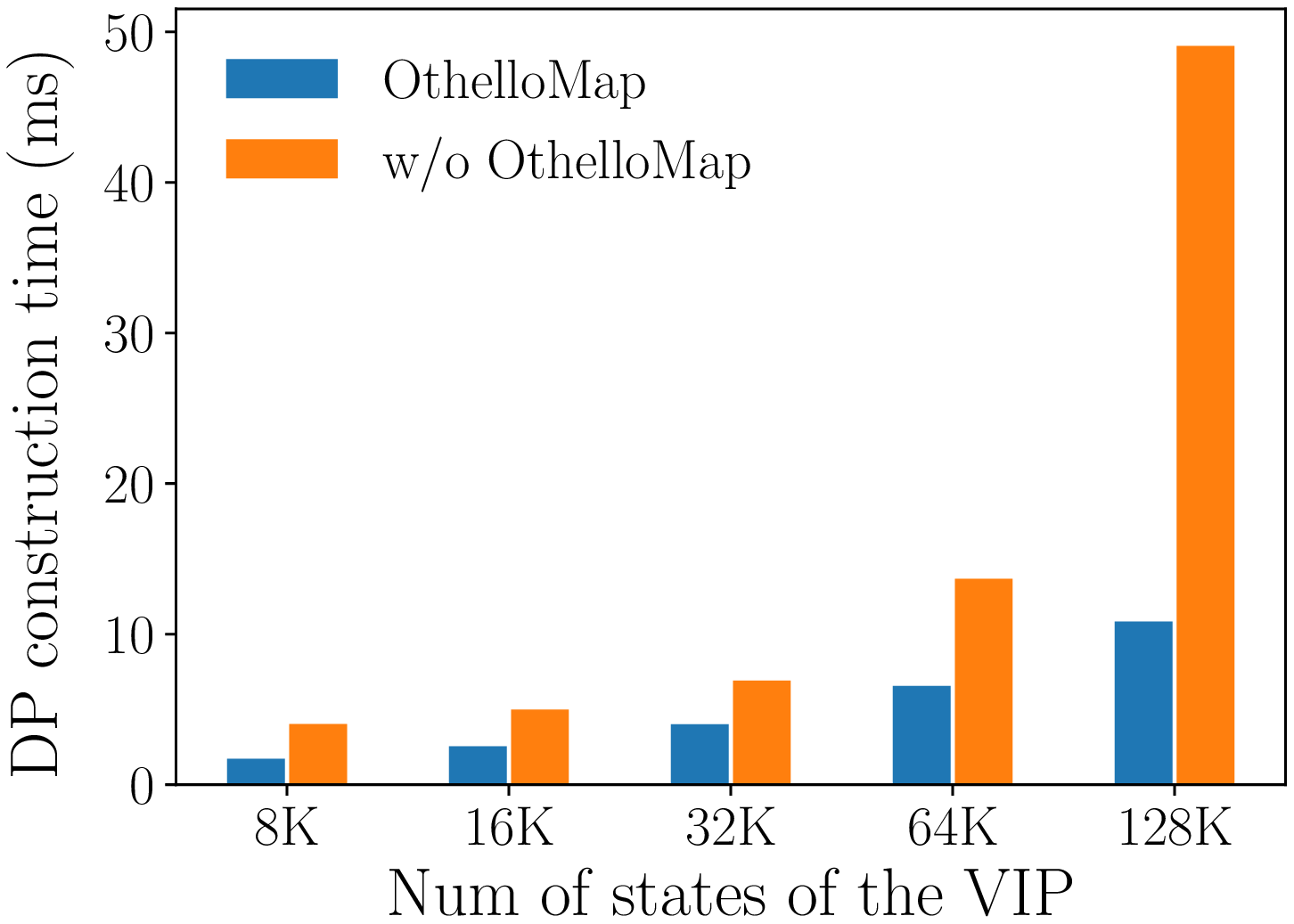}
    \caption{\small Response time of \sysn-DP construction}
    \label{fig:dp-update}
\end{tabular}
\end{figure*}

\subsection{Weighted load balancing}
\label{sec:weighted}

\textbf{Reason of using DIP code.}  One may notice that, to process the first packet of a new state, \sysn gets $Dcode$ and then transfers it to the DIP, rather than directly putting the DIPs as the lookup results of an Othello. It is because a DIP is 32-bit long, while a DIP code can be much shorter, e.g, 10 bits. Our method reduces the storage cost. The total number of distinct DIP codes, $2^l_d$,
can be larger than the number of DIPs, e.g., by about an order of magnitude, in order to provide the granularity for a weighted randomizer.
The $Dcode$ to $DIP$ mapping is determined by how the LB wants to assign the weights among DIPs of this VIP. For example, if $Dcode$ has 4 bits and there are 4 DIPs and all DIPs have equal weights, then we may map $Dcode$ in [0000, 0011] to $DIP_1$, $Dcode$ in [0100, 0111] to $DIP_2$, $Dcode$ in [1000, 1011] to $DIP_3$, and $Dcode$ in [1100, 1111] to $DIP_4$.  
We may consider $Dcode$ as a ball and each DIP as a bin.

We first show that for an unknown state, the probability that Othello will find a particular $Dcode$ is  uniformly distributed among all possible values of  $Dcode$.
Based on that, if the number of $Dcode$ is larger than the number of DIPs by a certain scale, we may use the $Dcode$-$DIP$ mapping to implement a weighted randomizer.

For a new state $c$, the lookup result of an Othello is
$Dcode=\tau(c)=A[h_a(c)]\oplus B[h_b(c)]$,
where $A[h_a(c)]$ and $B[h_b(c)]$ are both $l$-bit values. Assume that $A[h_a(c)]$ ($B[h_b(c)]$) has equal probability to be any element in array $A$ (array $B$), which is true if $h_a$ and $h_b$ are uniform hashes.  Each element in $A$ or $B$ can be either `determined' or `free'. A determined element corresponds to a white vertex as in the example of Figure~\ref{fig:OthelloConstL}, whose value should be fixed during the construction to provide correct lookups for current states. A `free' element corresponds to a grey vertex and its value is `not care'. We assign uniformly random values for every free element. As a result, if $A[h_a(c)]$ and $B[h_b(c)]$ are both determined, $Dcode$ is determined. If one of $A[h_a(c)]$ and $B[h_b(c)]$ is free, then $Dcode$ is random. We know that $A$ and $B$ both have $m$ elements and there are $m^2$ possible pairs of $A[h_a(c)]$ and $B[h_b(c)]$. Among them, only $n$ pairs produce determined values of $Dcode$ and the portion is $n/m^2 < 1/n$. Hence only a small portion of the results are determined and the others can be considered uniformly random. The overall results are not strictly uniform but are expected to be close to it.

We conduct a series of experiments to validate this uniformity. Figure \ref{fig:1024-bucket} shows one typical example. We let the value length $l=10$. Hence there are 1024 possible Dcodes. We enumerate all possible combinations of indexes of $A$ and $B$ and compute the resulting Dcodes.
The hash functions used in \sysn is CRC32.
We observe that using \sysn, the  combinations (stateless packets) are very evenly distributed to different Dcodes, with min, 10\%, mean, 90\%, and max values to be
925, 980, 1024, 1066, and 1120 respectively. The results of other experiments are similar.

We compare \sysn with MD5  and SHA256. Although MD5 and SHA256 are not strictly uniform, they are considered \emph{sufficiently uniform} and can be used for many applications such as digest computation. We show that \sysn is no worse than them in uniformity and is sufficiently good to use in practical systems.
We conduct two well-known  statistical tests, the chi-squared test and Kolmogorov-Smirnov test, to compare \sysn, MD5, and SHA256 with the uniform distribution. As shown in Figures \ref{fig:chi2-fail-bar} and \ref{fig:ks-fail-bar}, every of them fails around or less than 10\% of the tests, because they are not strictly uniform. \sysn is no worse than either MD5 or SHA256, especially when $l_d>11$ (Dcode count $>2048$). In our implementation we set $l_d=12$.
We will further evaluate the load distribution to DIPs in \S~\ref{sec:P4exp}.


\subsection{\sysn control plane}
\label{sec:OthelloMap}
The tasks of the \sysn control plane (\sysn-CP) are two-fold: 1) tracking existing states; and 2) generate new data plane structures, mainly the new Othello, when a data plane update is required.
A na\"{\i}ve  solution is to use a hash table to store a set of state-DIP pairs.
When an update is needed, the new Othello lookup structure is constructed from the set.
Our \textit{innovative idea} is to design a new data structure called OthelloMap that maintains both the state-DIP pairs and the Othello lookup structures for all current states. Note if the network includes $M$ VIPs, the control plane has $M$ OthelloMaps. The purpose of using OthelloMap is to quickly generate the new \sysn-DP when network dynamic happens.

\textbf{Components of an OthelloMap.}  An OthelloMap of VIP $v$ includes two parts.  1) An array $C$ of size $n$, where $n$ is the number of current states of VIP $v$. Each element of $C$ stores a state-DIP pair and the corresponding $Dcode$.  2) An Othello $O$ constructed using the set of current states. The lookup result of $O$, using the state identifier (ID) $c$, is the index $i$ such that $C[i]$ stores the state-DIP pair of $c$. Note the length of $i$ is no smaller than $\lceil\log_2n\rceil$ bits.

\textbf{Set query to OthelloMap.} Set query is a basic function of OthelloMap. The input is a possible state ID $c'$ and the output is either the corresponding DIP or `not exist'. To conduct a set query, the OthelloMap performs a lookup to the Othello $O$ using $c'$ and get a value $i$. If the state exists,  $C[i]$ includes the DIP. Otherwise, the connection stored in $C[i]$ does not match $c'$. Hence it can return `not exist'. This process takes $O(1)$ time.

\textbf{Addition/deletion to OthelloMap.} To add a state-DIP pair $\langle c, DIP\rangle$, to the OthelloMap, we first apply set query of $c$. If $c$ exists, $C[i]$ is revised to $\langle c, DIP \rangle$. If $c$ does not exist, we store $\langle c, DIP \rangle$ to $C[n+1]$. Then we add $\langle c, n+1\rangle$ to Othello $O$. This process takes average $O(1)$ time.
To delete a state-DIP pair $\langle c, DIP\rangle$ from the OthelloMap, we apply set query of $c$. If $c$ does not exist, we do nothing. Otherwise $c$ and its DIP are stored in $C[j]$. We delete them from $C[j]$ and the move the element in $C[n]$, say $\langle c', DIP' \rangle$, to $C[j]$. Then we revise the value corresponding to $c'$ in Othello $O$ from $n$ to $j$. This process  takes $O(1)$ time.

An illustration of OthelloMap is shown in \emph{Appendix D}.

\textbf{Memory cost analysis of \sysn-DP.} Let $l_i$ be the length of the index $i$ and $l_k$ be the length of each state-DIP pair information. The memory cost of \sysn-DP is $2.33l_in+ (l_k+l_d)n+ 64m+2^{l_d}l_vm+48*2^{l_v}$, where $2.33l_in$ is the overhead of the Othello $O$, $(l_k+l_d)n$ is the overhead of the array $C$, and the remaining is for the VIP array and DIP array that need to be updated to the data plane.

\textbf{Performance gain using OthelloMap.}
We compare the time to construct a new DP with and without OthelloMap. The results are shown in Fig.~\ref{fig:dp-update}.
\emph{OthelloMap significantly reduces the response time in the control plane during \sysn updates} by over 50\%.

\textbf{Interaction of \sysn-CP and Host Agents.} \sysn-CP receives state arrival/termination reports from Host Agents running on different DIP servers. Upon receiving a report, \sysn-CP performs corresponding addition/deletion operations to the corresponding OthelloMap.

We discuss Task 2 of \sysn-CP, i.e., how \sysn-CP generates new data plane structures for network updates in the next subsection.

\subsection{Reactive control/data plane update}

\sysn-CP does not have to update the \sysn-DP on receiving state arrival/termination reports. Instead, it only updates the \sysn-DP when there is a DIP-pool change. It is because only under a DIP-pool change, the current \sysn-DP may violate consistency.
Recall that \sysn-DP includes the VIP array, the Othellos for all VIPs, and the DIP array. For the change on a DIP pool of VIP $v_i$, only the Othello related to $v_i$ and the $i$-th dimension of the DIP array need to be updated, which are a relatively small portion of the entire \sysn-DP. All other parts can be kept still.

Updating the DIP array is based on the load balancing method introduced in \S~\ref{sec:weighted}, which is fast.
To generate the updated Othello  of $v_i$, denoted by $O_i$, we need to include all current states and remove terminated ones. The Othello of the OthelloMap of $v_i$, denoted by $O'_i$, includes all states. The only difference between $O_i$ and $O'_i$ is their lookup values ($Dcode$ versus OthelloMap index). Recall that the main computation complexity of Othello construction is to compute the acyclic  bipartite graph $G$ to include the set of keys. Once $G$ is determined, assigning the values of the keys is trivial, with complexity bounded by one-time pass of the values. Therefore we simply re-use the $G$ from the OthelloMap and assign the $Dcode$ values, which takes a short and bounded time.
At the end, \sysn-CP sends the updated structures to \sysn-DP using a programmable network API.

The pseudocode of \sysn-DP updating is presented in \emph{Appendix B}.
Upon receiving the update message, \sysn-DP only needs to modify the arrays related to one particular VIP. Since the memory spaces of all VIPs are independent, the modified memory size is very small (less than 1MB in most cases). The packets to other VIPs can be concurrently processed while updating the data plane. In addition, we design the \emph{concurrent control} method that locks 1024 bits at a same time for updating and only blocks packet lookups that need to access the 1024 bits.  Due to space limit we skip the details.

\textbf{Update complexity.} The time/space complexity of data plane update is in $O(l_dn_i)$, where $n_i$ is the number of connections of VIP $v_i$ and $l_d$ is the length of Dcode. Note \sysn update  happens infrequently, once per DIP change, and only applies to the part of data plane structures of one VIP.

\subsection{Consistency guarantee under dynamics}
\label{sec:dynamic}

An LB experiences three types of dynamics: 1) state arrival/termination; 2) DIP pool changes; 3) VIP changes.
It is important that packet consistency is still preserved during network dynamics.

For state arrival and termination, \sysn-DP has no change.
In this case every packet to a VIP $i$ will have three possibilities for the Othello lookup.
\begin{enumerate}
\item The state ID of the packet, $k$, is known by \sysn-CP during the construction of Othello-$i$, and the value of looking up $k$ is $Dcode$ which can be mapped to the DIP holding this state. Then the lookup result $\tau(k)=Dcode$ and the packet will be forwarded to the correct DIP.
\item The state ID $k$ is unknown by \sysn-CP  during the construction, and the packet is the first one of a  new state. Then according to the property of Othello, $\tau(k)$ is an arbitrary $l$-bit $Dcode$. According to the property of the table $DA$, $DA[i][Dcode]$  always stores a valid DIP for VIP $v_i$. Hence the packet will be forwarded to a valid DIP $D$.
\item The state ID $k$ is unknown by \sysn-CP  during the construction, and the packet is not the first one of a  new state. Hence the first packet was processed after the latest construction and update, which was forwarded to a DIP $D$. Since the data plane has not been updated since then, \sysn still returns $D$ as the DIP of this packet, which preserves consistency.
\end{enumerate}
\sysn does not cause false hits either. Using the three-level lookup structure, for any new TCP packet or UDP packet the corresponding Othello will return a  $Dcode$ that will be mapped to a valid DIP.

When a DIP pool change happens, the $Dcode$ to DIP mapping needs to be adjusted. Again using the example in Section~\ref{sec:dataplane}, we may map $Dcode$ in [0000, 0011] to $DIP_1$, $Dcode$ in [0100, 0111] to $DIP_2$, $Dcode$ in [1000, 1011] to $DIP_3$, and $Dcode$ in [1100, 1111] to $DIP_4$.
The state $c$ is mapped to 0100 and hosted on  $DIP_2$.
Suppose $DIP_4$ fails and the mapping is adjust as:
$Dcode$ in [0000, 0100] to $DIP_1$, $Dcode$ in [0101, 1001] to $DIP_2$, $Dcode$ in [1010, 1111] to $DIP_3$. Then the corresponding values in $DA$ should be adjust, e.g., $DA[i][0100]$ should be changed to $DIP_1$ from $DIP_2$. Also packets of the state $c$ should stick to $DIP_2$, hence we change its $Dcode$ to 0101 and revised the Othello structure accordingly. In this way, packet consistency is preserved.

VIP changes are very infrequent and can be dealt easily. It requires only adding an element to the VIP array and adding/deleting corresponding Othello and one dimension of the DIP array. No packet consistency is involved.

\emph{Therefore \sysn achieves packet consistency without requiring updating for every new state. It only updates when there is DIP change. This is a unique feature of \sysn compared to other stateful LBs to achieve processing and update efficiency.}

An additional mechanism is to prevent the possible consistency violation in the following situation: \sysn has started to update and then a packet of a new state comes in before \sysn finishes updating.  This is a common problem for all software LB designs.
Note compared to other methods that update in per-connection basis, \sysn updates in per-DIP-change basis, hence such problem happens very infrequently.

\section{Implementation and Evaluation}
\label{sec:evaluation}
In this section we evaluate the performance of the \sysn algorithm and prototype systems.
We implement two prototypes: 1) a software LB on commodity servers by Intel DPDK \cite{DPDK} deployed in CloudLab \cite{CloudLab}, a research infrastructure for cloud
computing experiments; and 2) a P4 \cite{P4} prototype running on Mininet \cite{mininet}. \textbf{Our code is available with an anonymous link \cite{ConcuryCode} and results can be reproduced.}



\begin{figure*}[t!]
\centering
\begin{tabular}{p{115pt}p{115pt}p{115pt}p{115pt}}
     \centering\includegraphics[width=\widthinTriColumn]{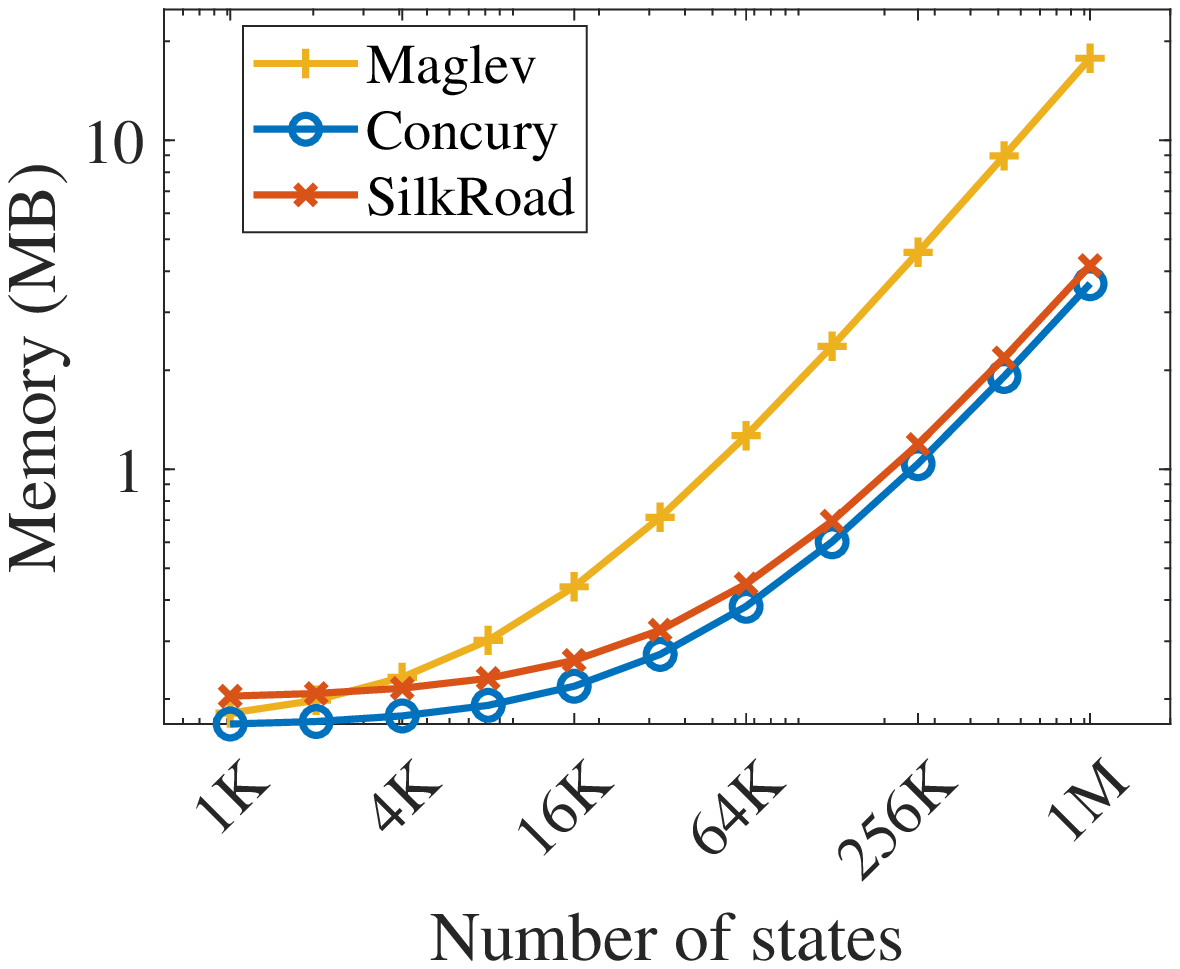}
     \caption{\small Memory cost for DIP-E and Small network}
    \label{fig:static-mem-sml}
    &\centering\includegraphics[width=\widthinTriColumn]{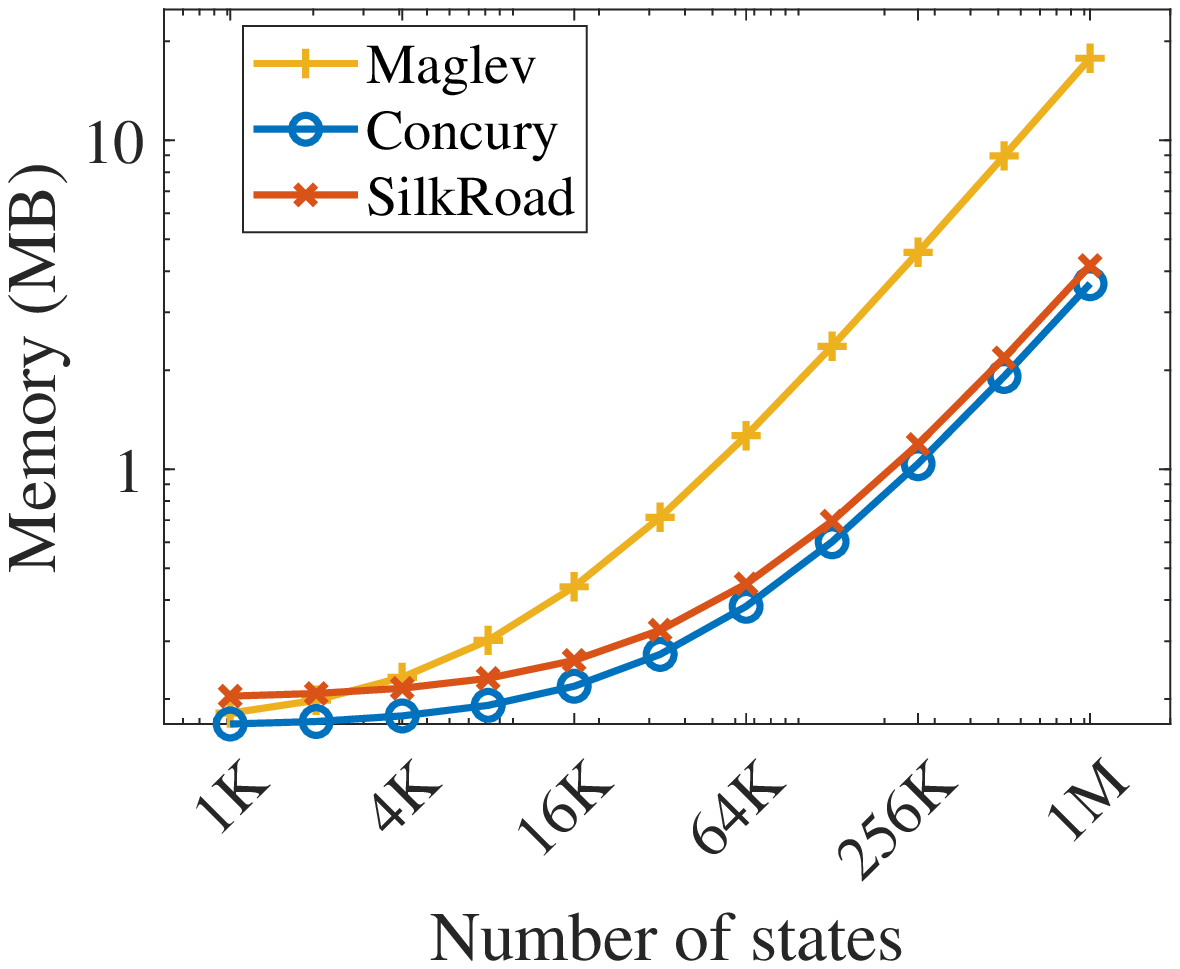}
    \caption{\small Memory cost for DIP-V and Small network}
    \label{fig:vary-dip-count-static-mem-sml}
    &
  \centering\includegraphics[width=\widthinTriColumn]{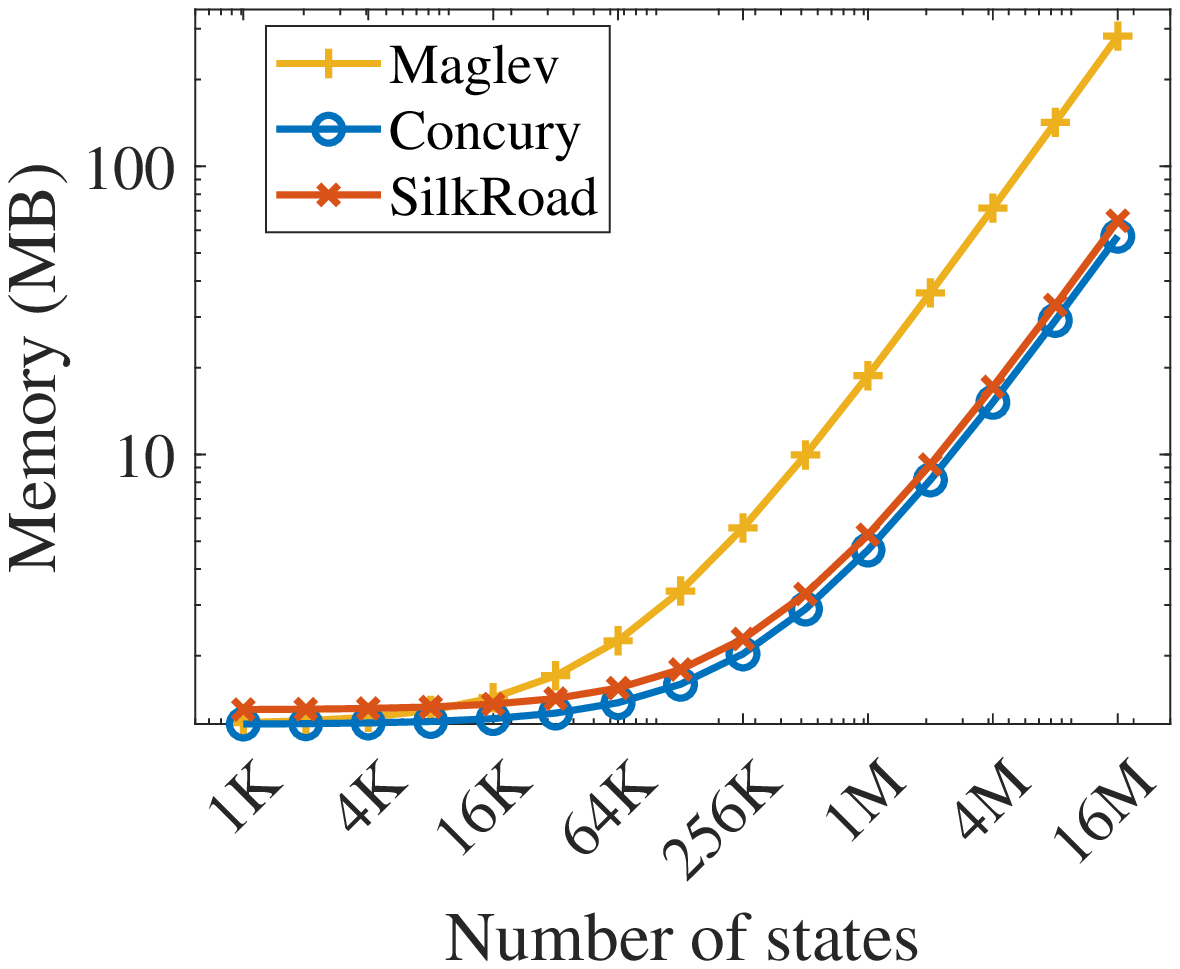}
     \caption{\small Memory cost for DIP-E and Large network}
    \label{fig:static-mem}
    &\centering\includegraphics[width=\widthinTriColumn]{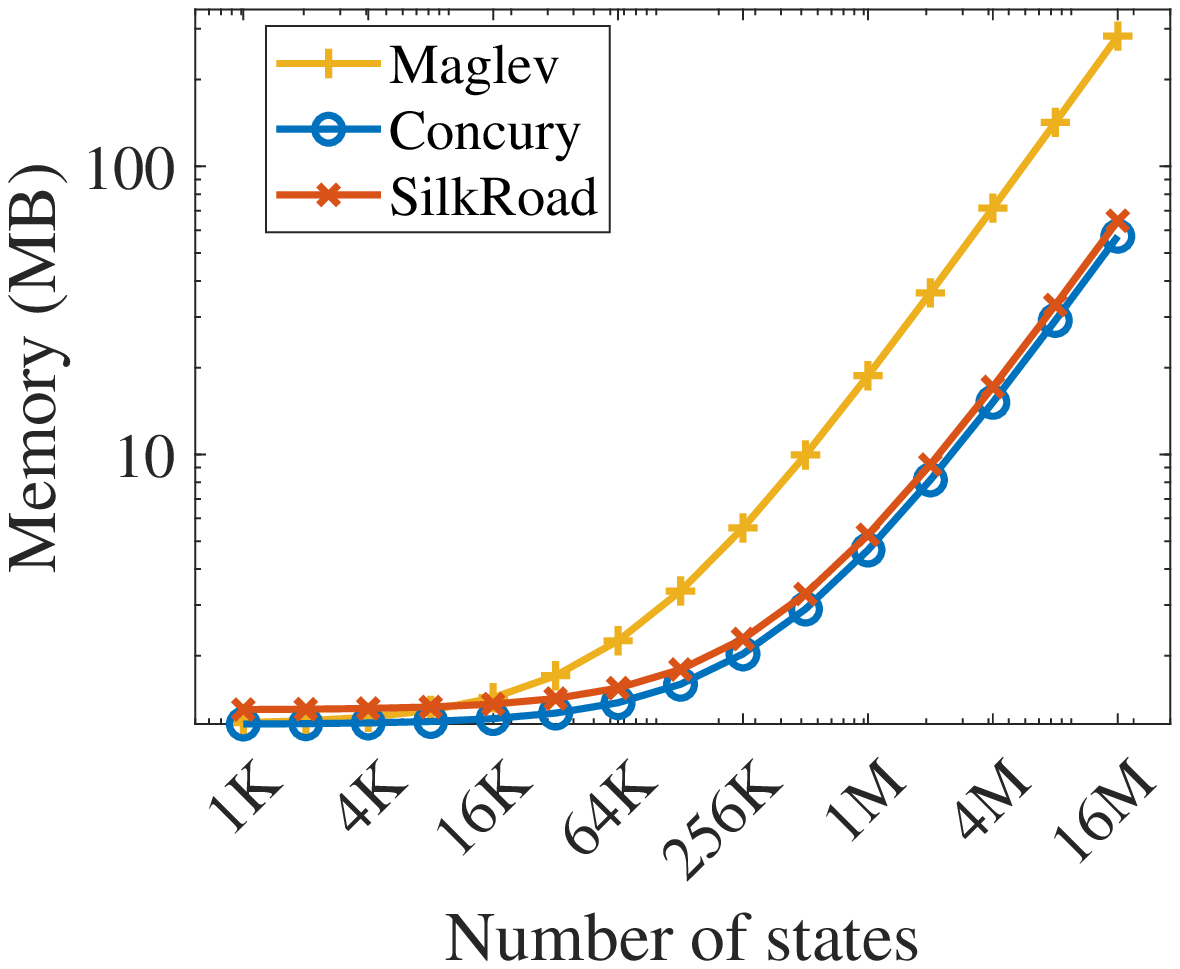}
    \caption{\small Memory cost for DIP-V and Large network}
    \label{fig:vary-dip-count-static-mem}
\end{tabular}
\end{figure*}


\subsection{Evaluation methodology}
We conduct three types of evaluations: 1) algorithm micro-benchmark; 2) software LB prototype using DPDK \cite{DPDK} deployed in CloudLab \cite{CloudLab}, and 3) a P4 prototype running on mininet \cite{mininet}.
The purpose of algorithm micro benchmark is to thoroughly compare the algorithm advantages of \sysn over existing solutions. The purpose of evaluating software LB in CloudLab is to show the actual performance of \sysn running in a real Cloud network. The purpose of the P4 evaluation is to show that \sysn can also be deployed to programmable switches.

We compare \sysn with two most-recent LB algorithms: 1) Hash table with digest, used in Maglev \cite{Maglev}; and 2) Multi hash tables with digest, used in SilkRoad \cite{SilkRoad}. Note SilkRoad was designed for special hardware, i.e., $>50$MB programmable switch ASICs. Hence, the performance shown in \cite{SilkRoad} is different.
We implement the LB algorithms of Maglev and SilkRoad \emph{in our best effort to improve their performance and ensure consistency}, but we are not able to re-build identical system prototypes of Maglev and SilkRoad as some of their technique details are not fully presented in \cite{Maglev} and \cite{SilkRoad}. In addition, we also separate the hash table of Maglev in a per-VIP basis --a fix to reduce potential digest collisions but not fully resolving it. 
We evaluate the performance metrics including memory cost, processing throughput, and load balancing.
For all experiments, we verify that packets of a same state are always sent to a same DIP.
\sysn causes neither packet consistency violation nor false hits.
The comparison of consistency violation and false hits would be just criticizing the other methods, hence we do not spend space to further show them. We do not compare \sysn with stateless LBs \cite{Olteanu2018,Araujo2018}. It is because stateless LBs use a relatively simple LB structure but move the overhead to server side. Hence it is hard to conduct a toe-to-toe comparison and the server upgrading cost is difficult to measure.

Note since Maglev is not open-source, we cannot provide identical performance as shown in the paper \cite{Maglev}. We use our best effort to improve their performance and ensure consistency. Also the performance depends on the computing platform of running these algorithms.

We use CRC32-C \cite{Intelcrc} for robust and faster hash results in \sysn. Recall that the construction of Othello may need sufficient different hash functions. We generate these hash functions using the following approach. Let $H$ be a CRC32 hashing and  $\mathtt{seed}$ be a 32-bit integer. We let $h_a(k) = H(k,\mathtt{seed}_a)$ and $h_b(k) = H(k,\mathtt{seed}_b)$.
Thus, $h_a$ and $h_b$ are uniquely determined by  $\mathtt{seed}_a$ and $\mathtt{seed}_b$ respectively.

We use the \textbf{real  traffic trace} from the Facebook data center networks \cite{FBtrace} for experiments. Since the packets in the trace only carry the DIPs, we assign them to 128 VIPs. We also generate synthetic traffic for production runs and dynamic experiments over a duration of time. We generate two setups of synthetic traffic: 1) \emph{DIP-E}. All VIPs have the same number of DIPs, and they have the same number of concurrent states at any time.  2) \emph{DIP-V}. VIPs have varied numbers of DIPs, and the numbers of concurrent states also vary with the numbers DIPs.
The number of VIPs may be 128 or 256.
We also consider two types of networks:
The \emph{Small} network models an Edge and the \emph{Large} network models a cloud. In the Small network, each VIP has 32 DIPs for DIP-E and 8 to 64 DIPs for DIP-V (32 in average).  In the Large network, each VIP has 128 DIPs for DIP-E and 32 to 256 DIPs for DIP-V (128 in average).
We vary the number of states from 1K to 16M for Large and 1K to 1M for Small, which covers the range of practical networks.
According to actual measurement \cite{SilkRoad}, the 99th percentile number of concurrent connections in the PoP cluster of a large web service provider is smaller than 10M. Other types of clusters and edge networks have fewer active states, varying from a few thousands to 10M.

For most experiments we conduct production runs for at least 20 times and take the average. The variations are little and difficult to show in the figures.

\begin{figure*}[t!]
\centering
\begin{tabular}{p{115pt}p{115pt}p{115pt}p{115pt}}

    \centering\includegraphics[width=\widthinTriColumn]{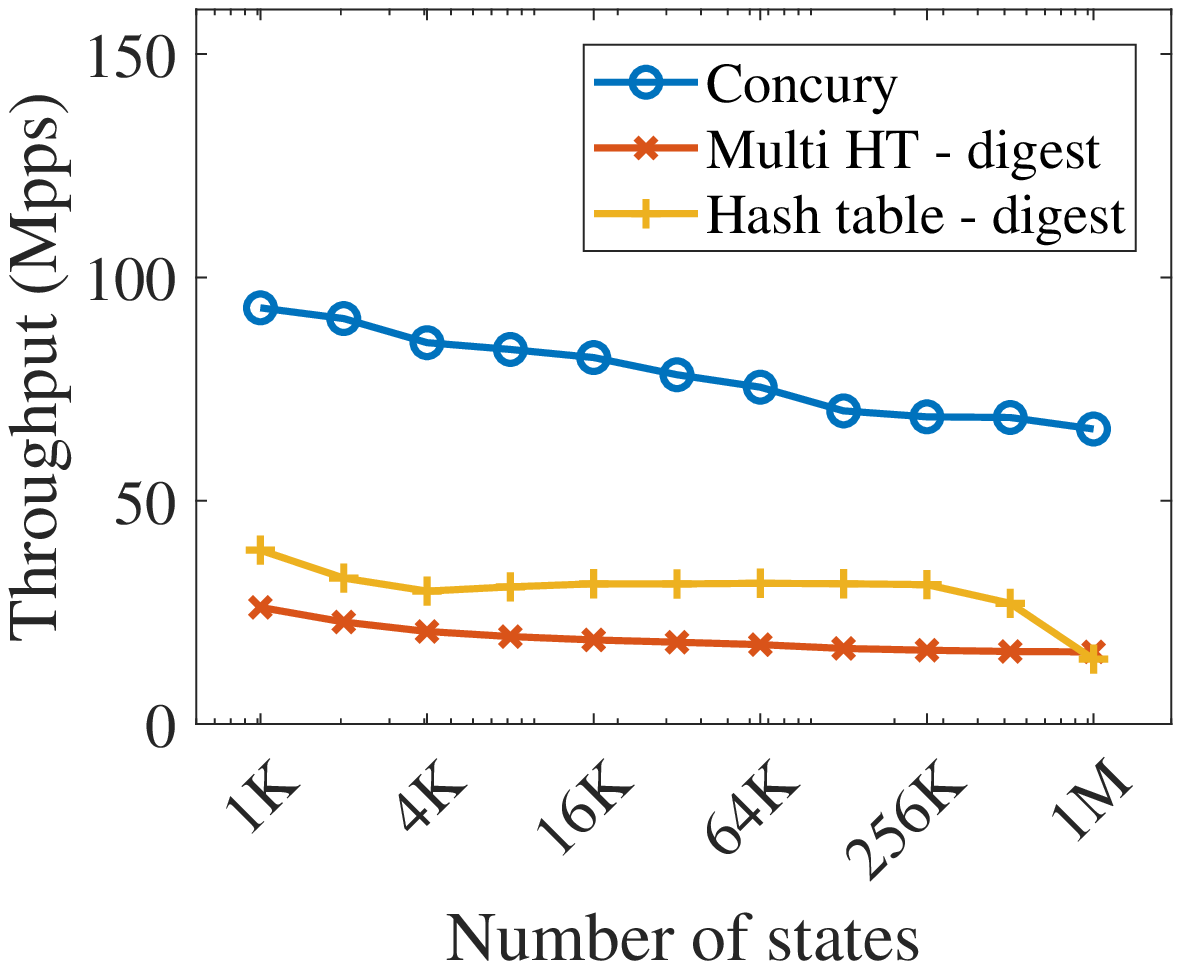}
    \caption{\small Throughput for DIP-E and Small network}
    \label{fig:static-query-sml}
    &\centering\includegraphics[width=\widthinTriColumn]{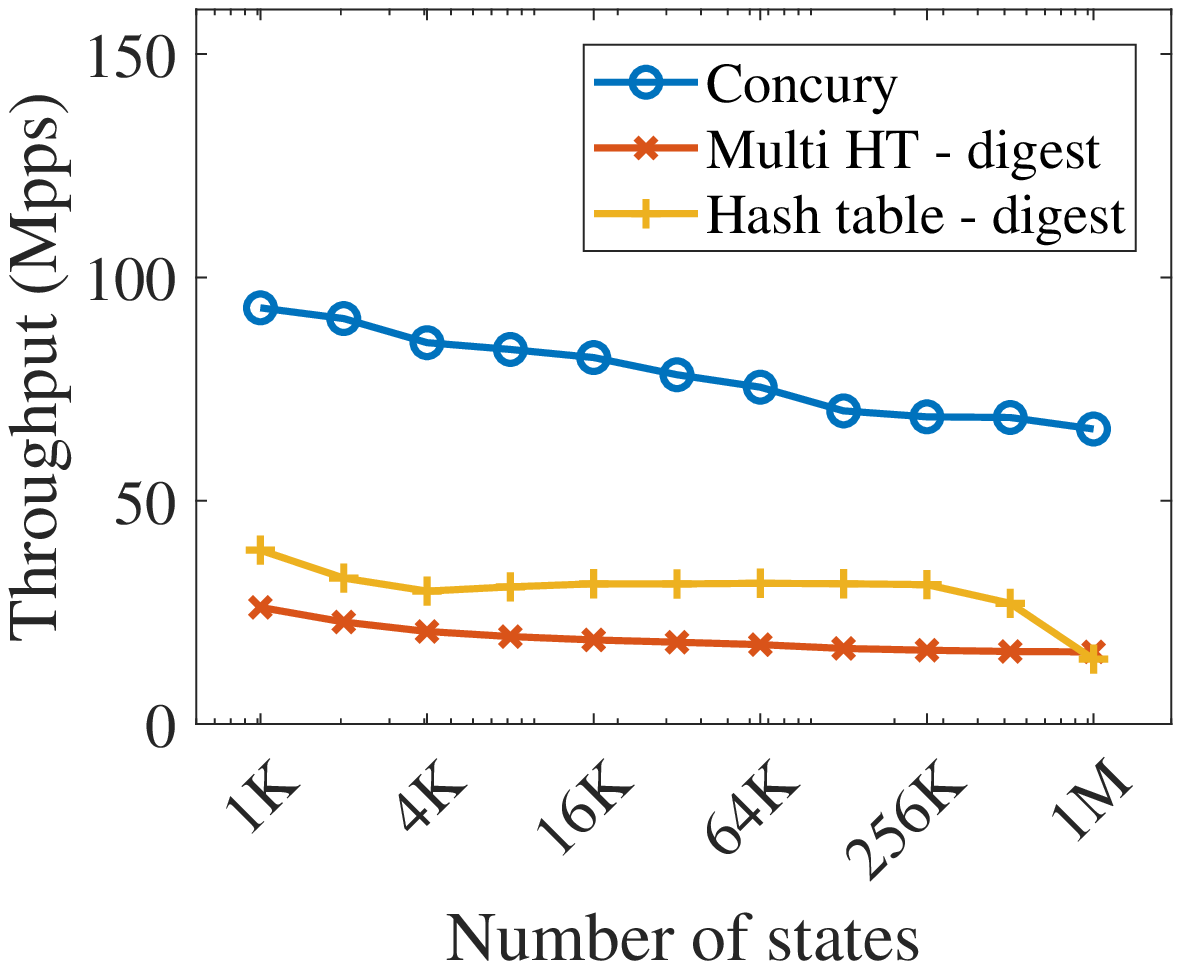}
    \caption{\small Throughput for DIP-V and Small network}
    \label{fig:vary-dip-count-static-query-sml}
    &\centering\includegraphics[width=\widthinTriColumn]{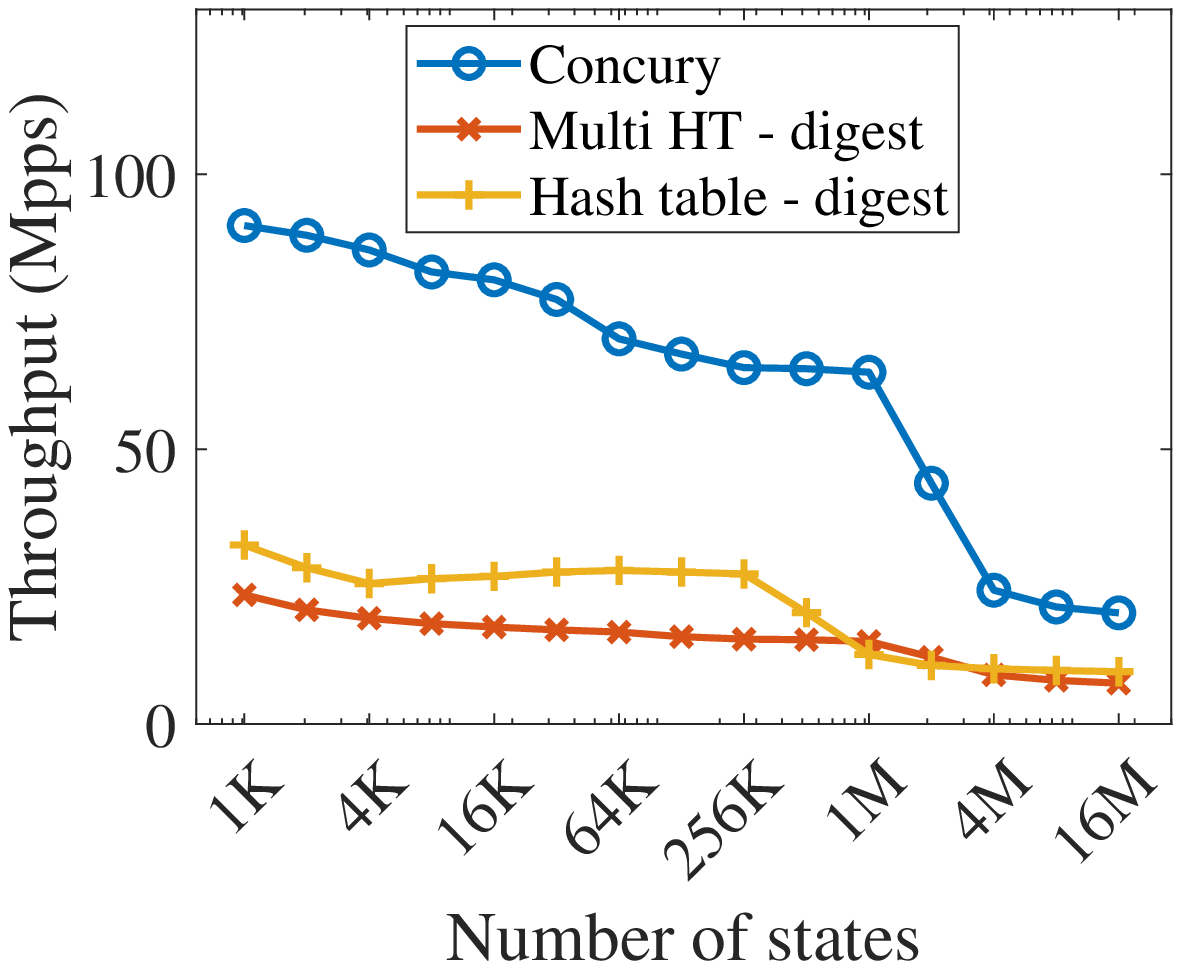}
    \caption{\small Throughput for DIP-E and Large network}
    \label{fig:static-query}
    &\centering\includegraphics[width=\widthinTriColumn]{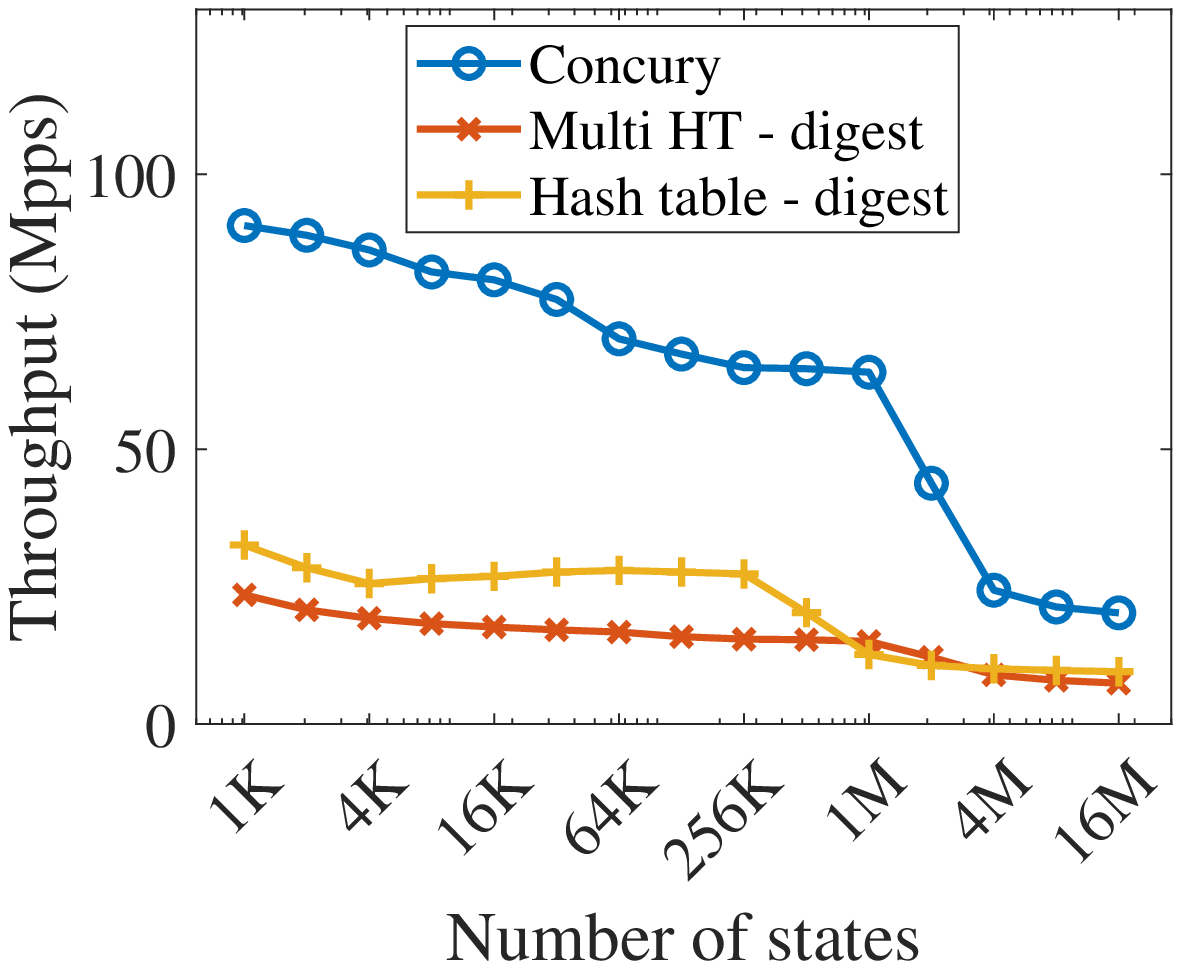}
    \caption{\small Throughput for DIP-V and Large network}
    \label{fig:vary-dip-count-static-query}
\end{tabular}
\end{figure*}

\begin{figure}[t!]
\centering
\begin{tabular}{p{115pt}p{120pt}}
\centering\includegraphics[width=\widthinTriColumn]{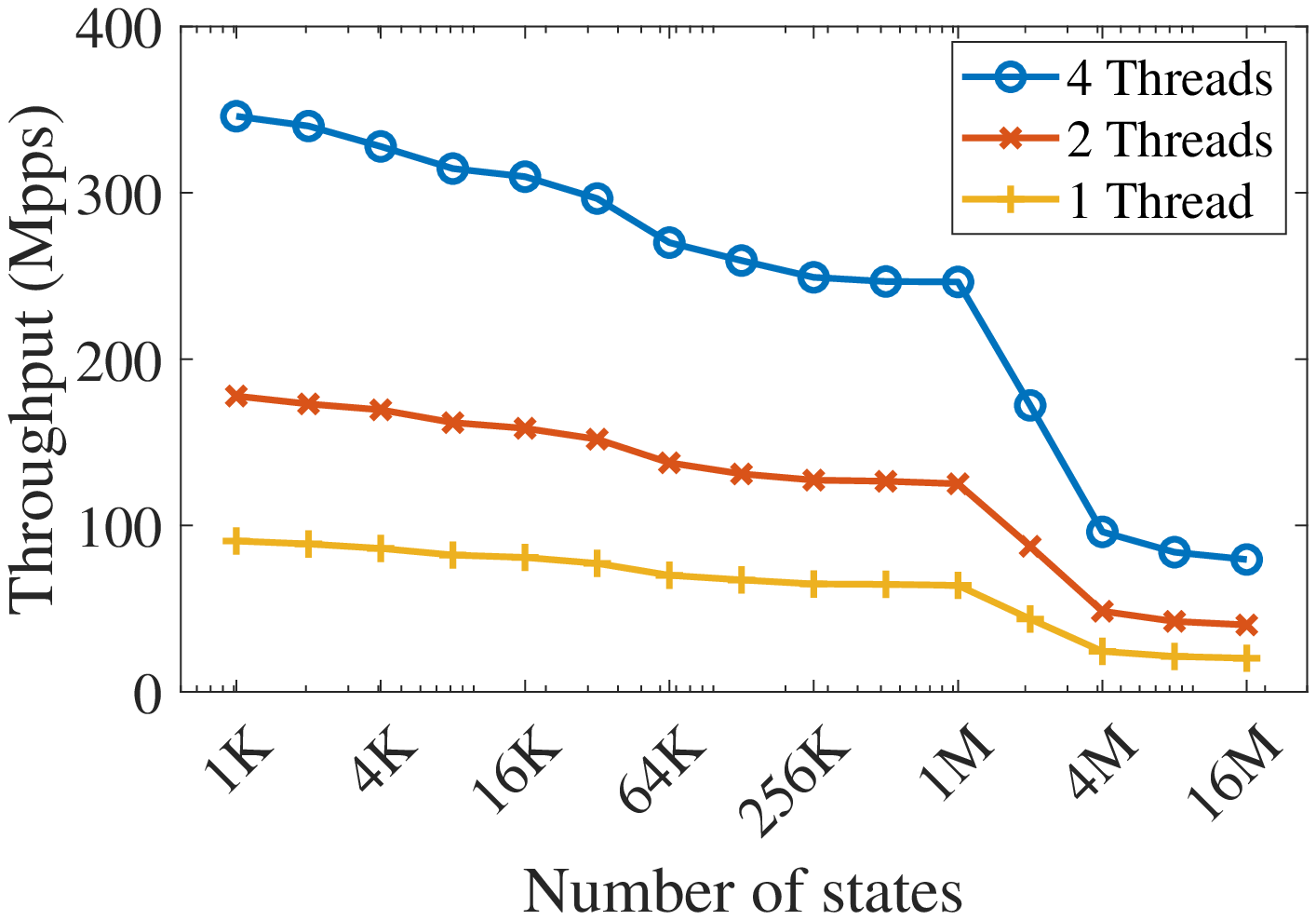}
    \caption{\small Throughput for multi-thread}
    \label{fig:multithread-throughtput}
&
\includegraphics[width=\linewidth]{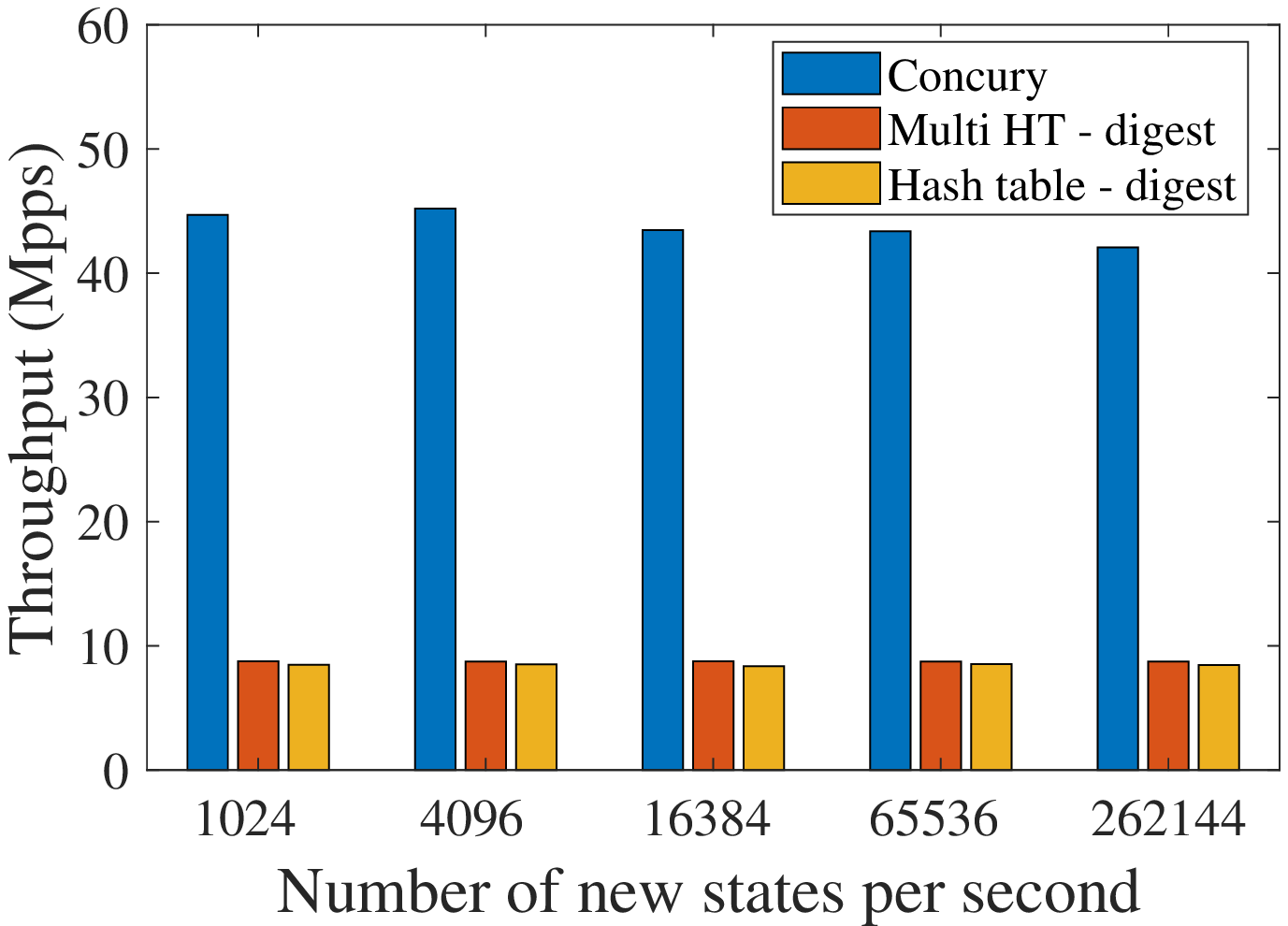}
    \caption{\small Throughput during data plane updates}
    \label{fig:dynamic-throughput}
\end{tabular}
\end{figure}

\subsection{LB algorithm evaluation}
\label{sec:commodity}

\textbf{Algorithm implementation details.}
We have implemented the complete functions of both \sysn-DP and \sysn-CP on a commodity desktop server with Intel i7-6700 CPU, 3.4GHZ, 8 MB L3 Cache shared by 8 logical cores, and 16 GB memory (2133MHz DDR4).
Different components of \sysn are interacted as in Figure~\ref{fig:ConcuryDP}. In addition, we need to provide a series of packets from different states and let \sysn process them.
One straightforward approach is to feed the LB with an existing traffic trace.
However, the time for transmitting the data from the physical memory to the cache is too long compared to the packet processing time on \sysn.
Hence, we use a linear feedback shift register (LFSR) to generate the states (identified by the 4-tuple) of every packet. The generated states are uniformly distributed over all possible 4-tuples, which is the worst case for load balancing performance for the lack of time locality.
One LFSR generates about 200M names per second on our server.
In addition, we provide event-based simulation using real traffic data to study the processing delay on \sysn.
Note that LFSR gives no favor to \sysn  because the names are generated in a round-robin scenario, which provides the minimum cache hit ratio.
LFSR traffic is actually the \textit{worst} traffic for \sysn compared to practical situations.
We use 1883 lines of  C++ code in total for this prototype.

\textbf{Memory efficiency.}
Figures~\ref{fig:static-mem-sml} and \ref{fig:vary-dip-count-static-mem-sml} show the memory cost of the LB algorithms of \sysn, Maglev, and SilkRoad in Small networks for the DIP-E and DIP-V setups respectively. We find that the memory cost of \sysn is less than 1MB for $<$256K states and 4MB for 1M states. The memory is only 20\%-30\% of that of Maglev, when the number of  states is $>$64K. It is very close to that of SilkRoad.
We also show the memory cost results in Large networks in Figures~\ref{fig:static-mem} and \ref{fig:vary-dip-count-static-mem}. \sysn has similar advantages compared to Maglev.
When there are 8M concurrent states, both \sysn and SilkRoad use $<38$MB.
The memory cost for  the DIP-E and DIP-V setups has no big difference.
\sysn is very efficient in terms of memory cost: it can be implemented on hardware switches with limited programmable ASICs or commodity servers that have limited cache.
Both Maglev and SilkRoad use digests, which introduce false hits.  \sysn provides false-hit freedom using similar or less memory.

\textbf{Processing throughput.}
The processing throughput of an LB algorithm characterizes its capacity. With a higher throughput, the network needs to deploy fewer instances of the LB and the infrastructure cost is reduced. Figures~\ref{fig:static-query-sml} and \ref{fig:vary-dip-count-static-query-sml} show the throughput of the LB algorithms of \sysn, Maglev, and SilkRoad in Small networks, using \textbf{a single thread} on a commodity desktop, for the DIP-E and DIP-V setups respectively. The metric is in millions of packets per second (Mpps).
Note SilkRoad was designed for programmable switch ASICs.
We implement the algorithm used in SilkRoad, named `Multi-level Hash Tables with Digest' (Multi HT-digest), on commodity servers and compared it to \sysn. Similarly, we also implement the algorithm used in Maglev, named `Hash Table with Digest'.
\sysn achieves $>65$Mpps when the number of concurrent states is $<1M$ and shows $>2$x advantage compared to Hash Table with Digest and Multi HT-digest.
For Large network results shown in Figures~\ref{fig:static-query} and \ref{fig:vary-dip-count-static-query},
when the number of states is $>1M$, the throughput reduces because the memory size is larger than the CPU cache size. However, \sysn still maintains the $>2$x advantage in throughput. The main reason resulting in the throughput advantage of \sysn is that the data plane of \sysn requires simpler operations than others. In addition, we show the throughput of \sysn for multiple threads in Fig.~\ref{fig:multithread-throughtput}. The result shows the throughput scales well with the number of threads: it reaches $>250$Mpps with $<1M$ states. The threads share the same memory space and do not compete for cache space.

\textbf{Cost of data plane update.}  Data plane updates consume CPU time. Hence, on a single thread, if data plane updates are complex, the throughput will evidently downgrade.
Existing LBs have no concurrent read/write designs \cite{Maglev,SilkRoad}.
We conduct the following set of experiments to evaluate the impact of updates to \sysn-DIP performance. We set the number of concurrent states to 1M and let new states join the network. The arrival rate ranges from 1K per second to 256K per second, reflecting the arrival rate in real networks. The DIP pools also change once per 10 seconds. The throughput during updates is shown in Fig.~\ref{fig:dynamic-throughput}. The throughput of Hash table-digest (in Maglev) and Multi HT-digest (in Silkroad) clearly downgrades (to $<$10Mpps) compared to the results shown in the static experiments in Fig.~\ref{fig:vary-dip-count-static-query}. \sysn experiences downgrading too (to 42Mpps), but the impact is limited. Hence, the data plane update cost of \sysn is small compared to other methods.

\begin{figure*}[t!]
\centering
\begin{tabular}{p{116pt}p{116pt}p{116pt}p{116pt}}
    \centering\includegraphics[width=0.95\linewidth]{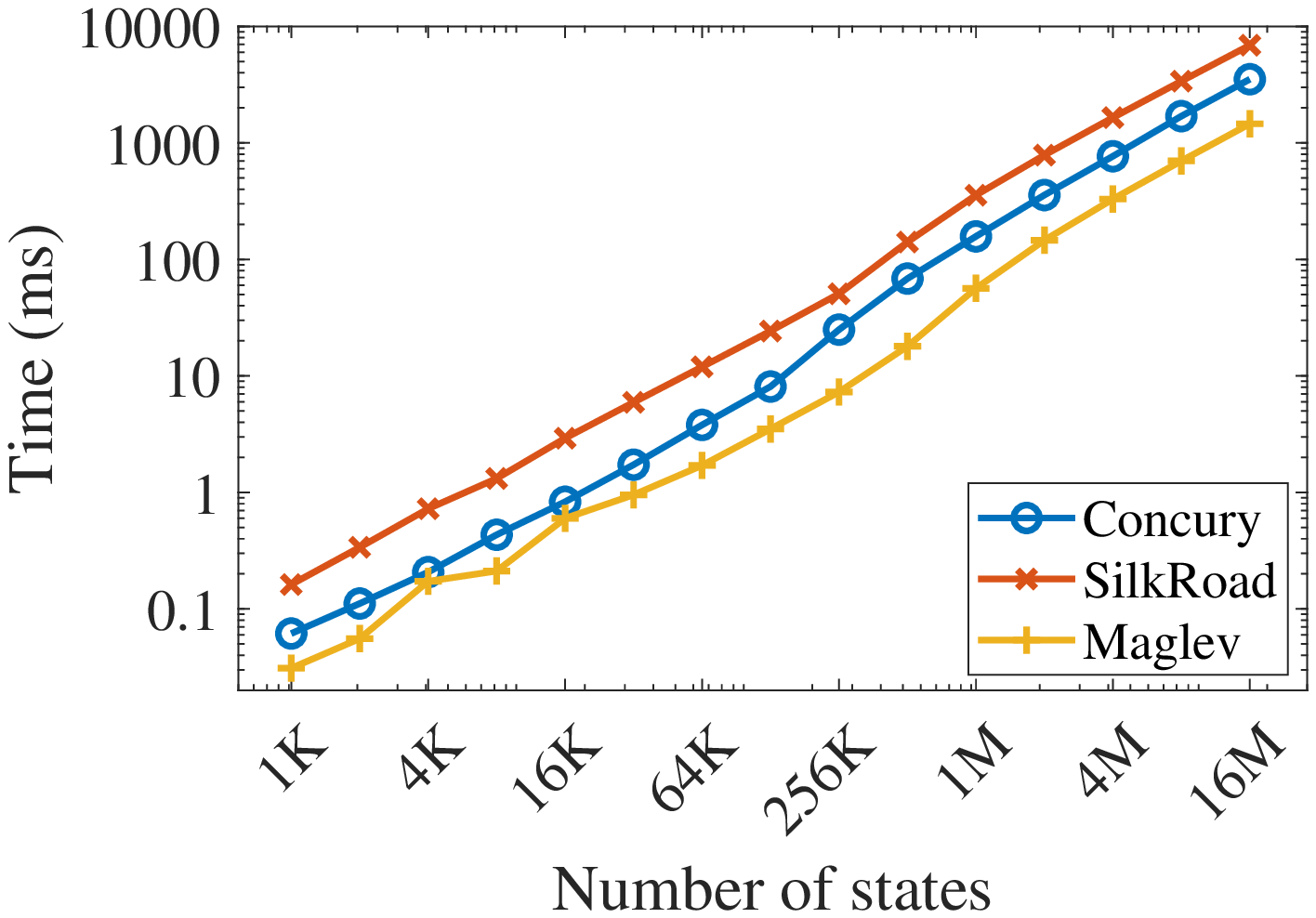}
    \label{fig:vary-dip-count-static-add}
    &\centering\includegraphics[width=0.95\linewidth]{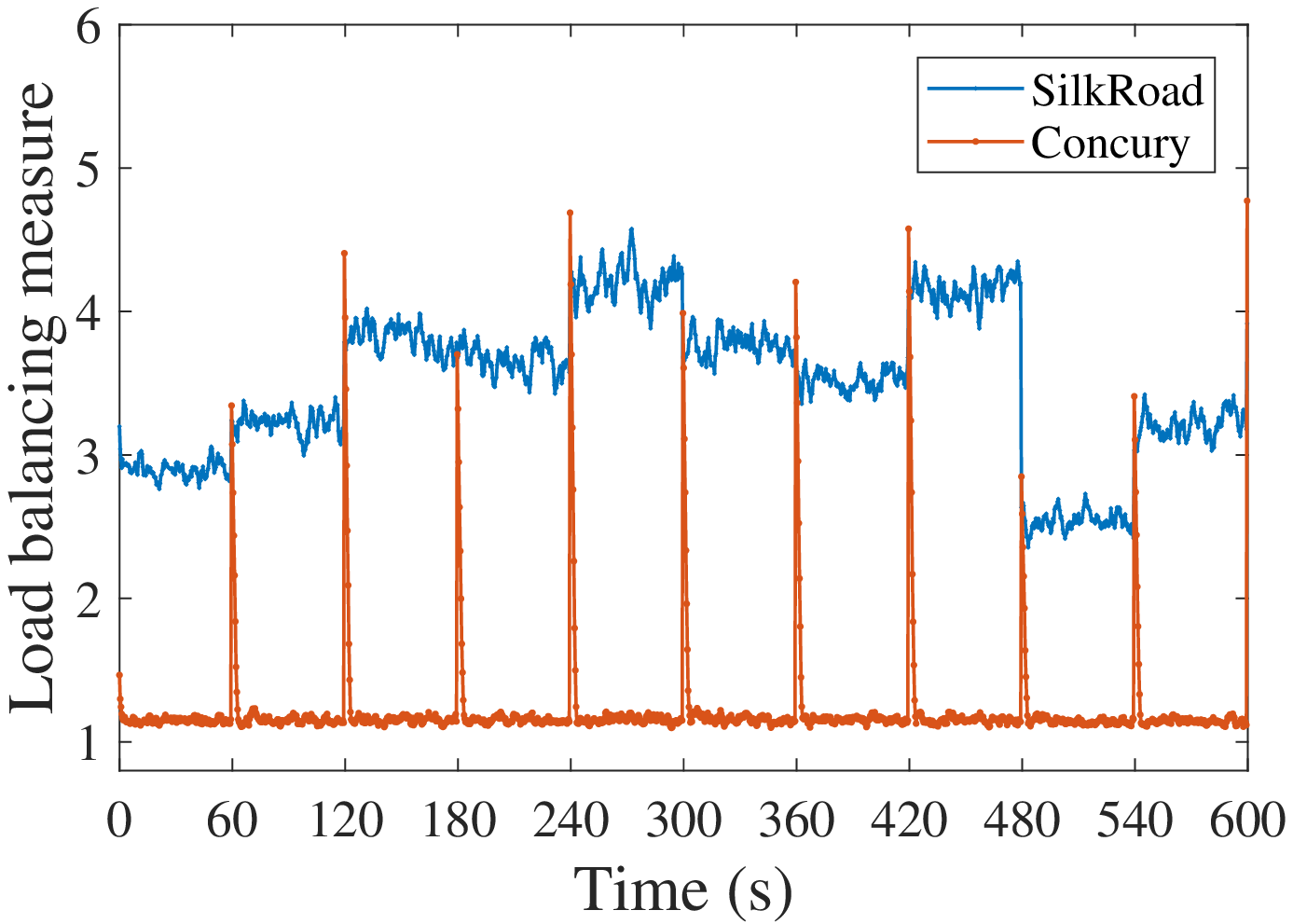}
    \label{fig:dynamic-load-balance-60}
    &\centering\includegraphics[width=0.95\linewidth]{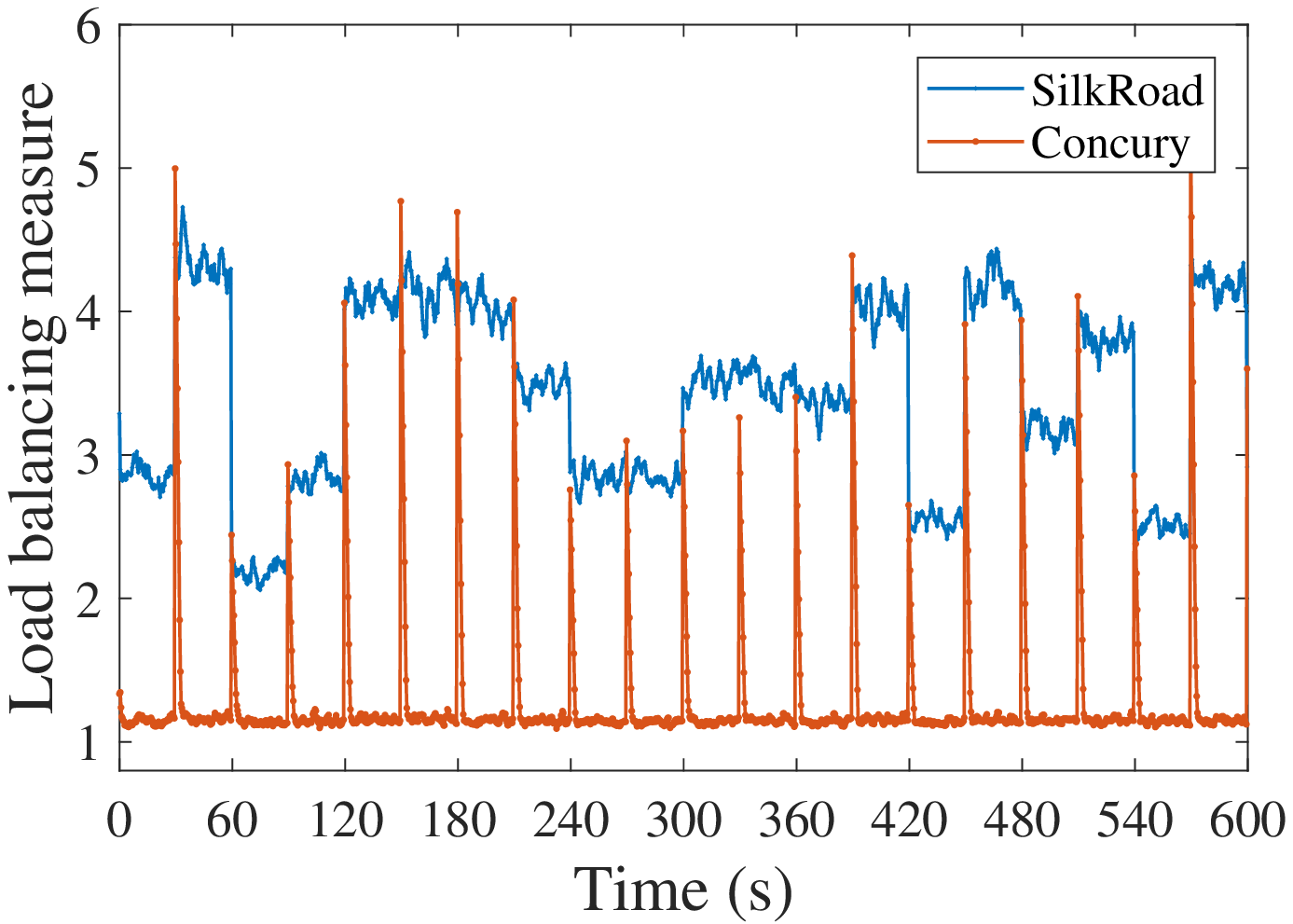}
    \label{fig:dynamic-load-balance}
    &\centering\includegraphics[width=0.95\linewidth]{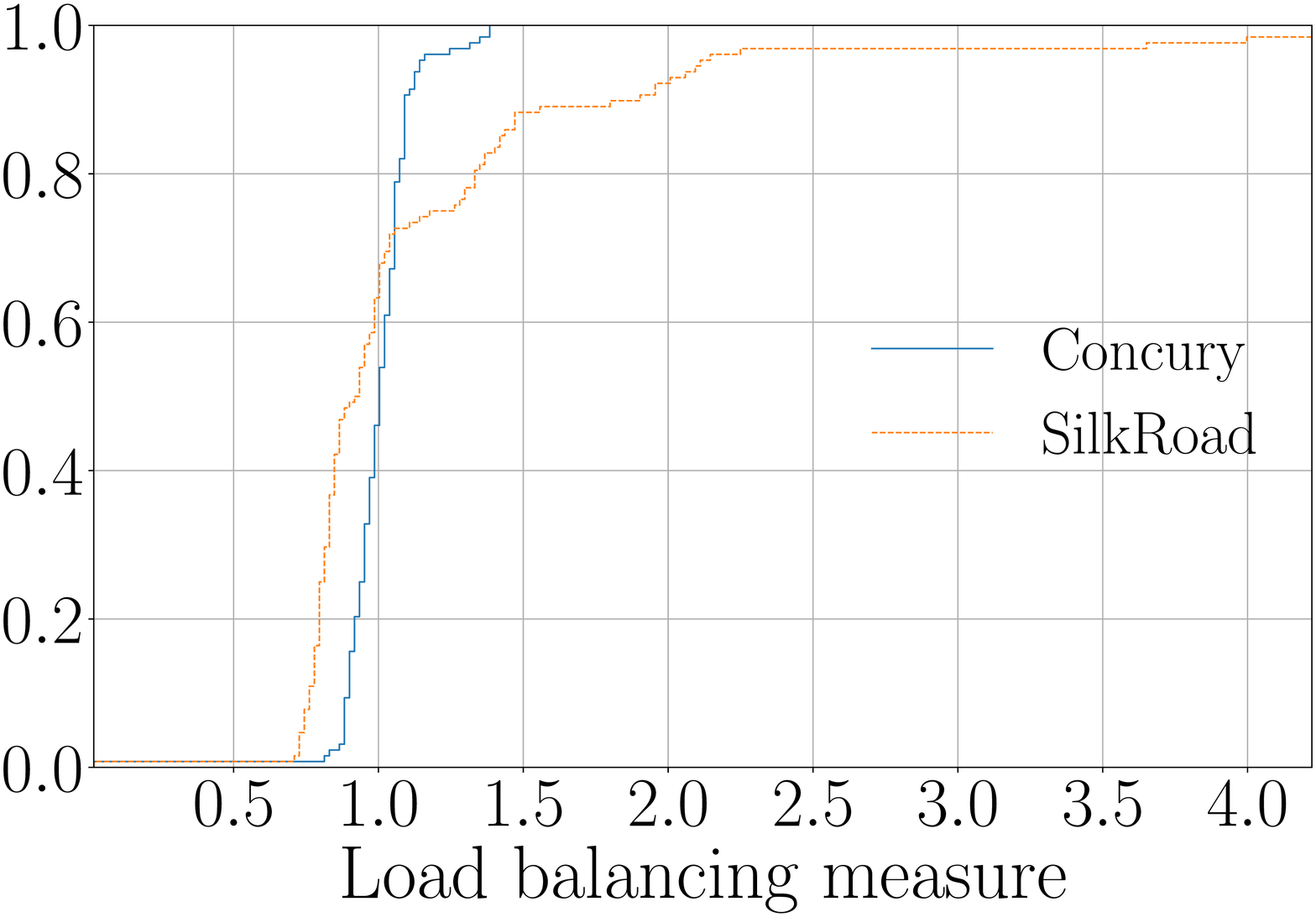}
    \label{fig:concury-silkroad-balance-measure-cdf}
\end{tabular}
\end{figure*}

\begin{figure*}[t!]
\centering
\begin{tabular}{p{160pt}p{160pt}p{160pt}}
    \centering
    \includegraphics[width=0.95\linewidth]{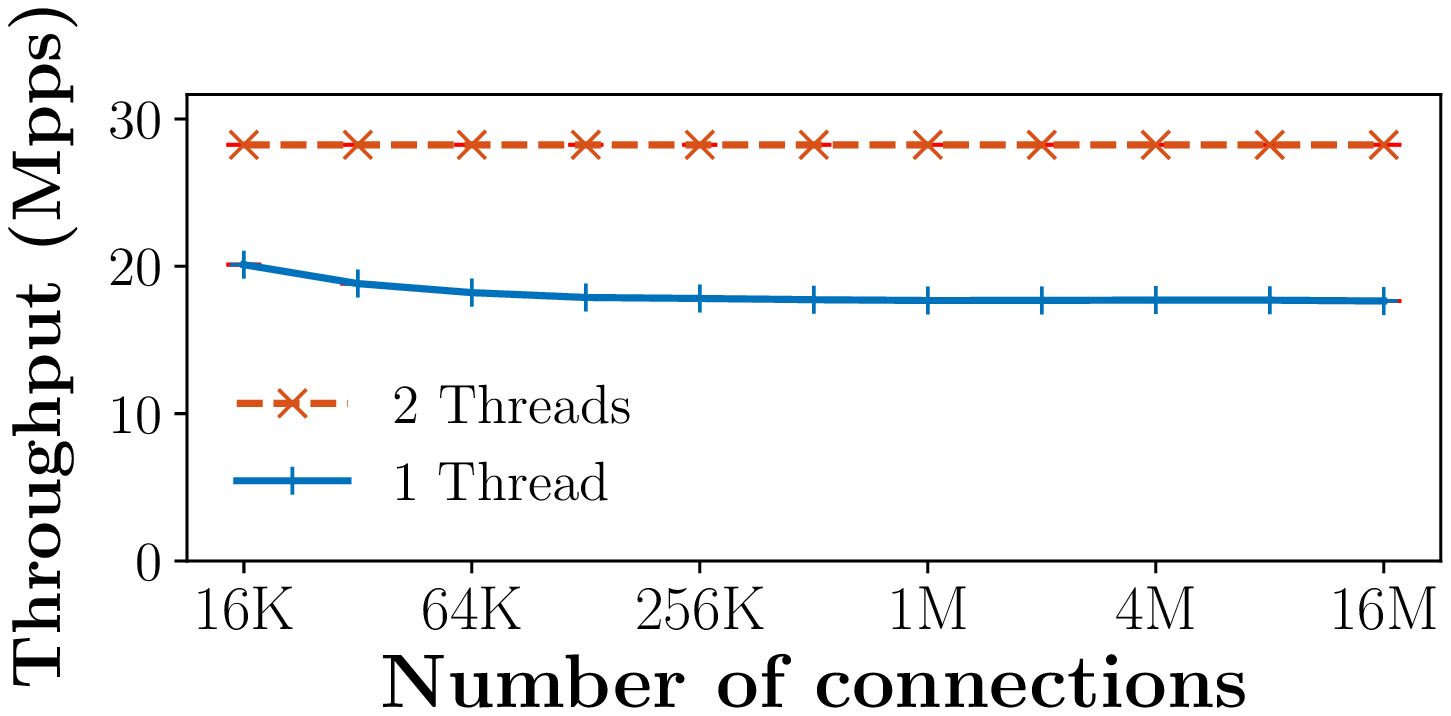}
    \caption{\small Throughput for DIP-V traffic by DPDK}
    \label{fig:dpdk-concury}
&
    \centering
    \includegraphics[width=0.95\linewidth]{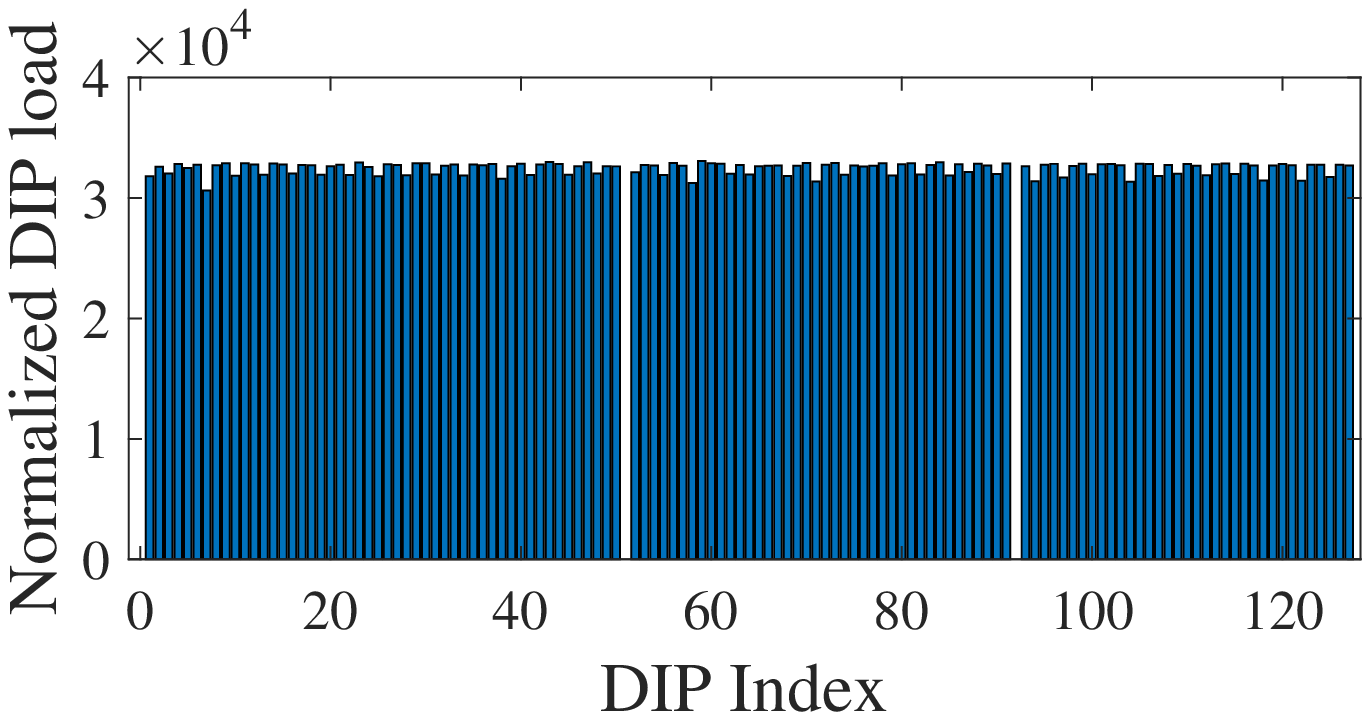}
    \caption{\small Normalized DIP load by P4 (real traffic)}
    \label{fig:REALConnoverWeight}
&
    \centering
    \includegraphics[width=0.95\linewidth]{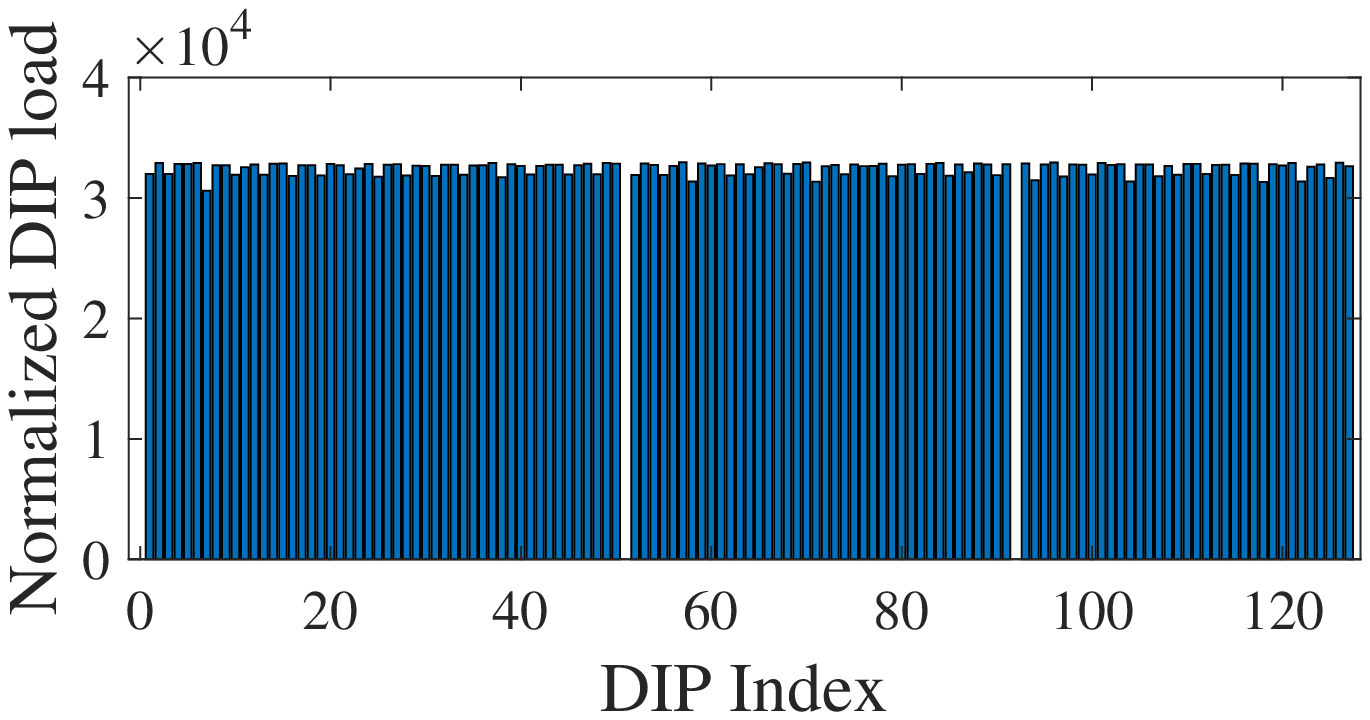}
    \caption{\small Normalized DIP load by P4 (synthetic traffic)}
    \label{fig:LFSRConnoverWeight}
\end{tabular}
\end{figure*}

\textbf{Response time and scalability of Control plane update.}  We show the performance of \sysn-CP in two aspects: 1) Response time of a DIP/weight change; and 2) Update time for new states. When a DIP/weight change happens, both the control and data planes need to be updated to reflect the change. \sysn-DP provides a tremendous advantage in the response time because of the invention of OthelloMap as shown in Fig.~\ref{fig:dp-update}. We find that when there are 8K to 128K states for one VIP (1M to 16M in total), the Concury-CP response time is only 2-12ms. On the other hand, Maglev requires very complex updates because it uses digests rather than the entire keys in the hash table. We further show the time cost of inserting new states to the control plane in Fig.~\ref{fig:vary-dip-count-static-add}. Note both the $x$ and $y$ axes are in logarithmic scale, and all three curves increase linearly with the number of the new states. For 16M new states, it only takes \sysn a few seconds to complete all updates. 
Hence, \sysn-DP is sufficiently fast and scalable to complete updates.

\subsection{Evaluation of software LB in CloudLab}
\label{sec:DPDKexp}

\textbf{Implementation details.} We implement \sysn as a software LB using Intel  Data Plane
Development Kit (DPDK) \cite{DPDK} running in CloudLab \cite{CloudLab}.
DPDK is a series of libraries for fast user-space packet processing \cite{DPDK}. DPDK is useful for bypassing the complex networking stack in Linux kernel and it has utility functions for huge-page memory allocation and lockless FIFO, etc. We modified the code of \sysn-DP and link it with DPDK libraries.
CloudLab \cite{CloudLab} is a research infrastructure to host cloud computing experiments. Different kinds of commodity servers are available from its 7 clusters. We use two nodes c220g2-011307 (Node 1) and c220g2-011311 (Node 2) in CloudLab to construct the evaluation platform of \sysn software LB prototype. Each of the two nodes is equipped with one Dual-port Intel X520 10Gbps NIC, with 8 lanes of PCIe V3.0 connections between the CPU and the NIC. Each node has two Intel E5-2660 v3 10-core CPUs at 2.60 GHz. The Ethernet connection between the two nodes is 2x10Gbps. The switches between the two nodes support OpenFlow \cite{OpenFlow} and are claimed to provide full bandwidth. 

Logically, Node 1 works as both a series of clients and a number of backend servers (DIPs) in the cloud, and Node 2 works as the software LB.
Node 1 uses the DPDK official packet generator Pktgen-DPDK \cite{Pktgen} to generate random packets and sends them to Node 2. The packets sent from Node1 carry the destination VIPs uniformly sampled from a set of valid VIPs. \sysn is deployed on Node 2 and forwards each packet back to Node 1 after determining and rewriting the DIP of the packet.
The packets will be forwarded by the \sysn node (Node 2) and  carry actual DIPs. By specifying a virtual link between the two servers, CloudLab configures the OpenFlow switches such that all packets from the \sysn node, with different destination DIPs, will be received by Node 1. Node 1 then checks the packet consistency to DIPs and records the receiving bandwidth as the throughput of the whole system.


In the real network, the results show that the \sysn software LB achieves 100\% packet consistency and the load balancing results are identical to those in the algorithm evaluation.
Fig.~\ref{fig:dpdk-concury} shows the throughput of \sysn for DIP-V traffic in CloudLab, measured Mpps, where all packets are 64 byte in length. We repeat 10 times for each data point, and the variations are too small to show as the error bars in the figure.
We first evaluate the maximum capacity of the 2x10GbE NICs by a simple forwarder that reads the 5-tuple of each packet and transmits it to the incoming port without looking up any FIB or table. The maximum capacity is 28.24 Mpps.\footnote{28.24 Mpps equals to 14.5 Gbps. We find it is common that the maximum transmission capacity is less than the NIC bandwidth. For example, Maglev \cite{Maglev} deployed by Google shows that its maximum capacity on a 10GbE NIC is 12 Mpps (=6.14 Gbps).} We evaluate up to 16M concurrent connections in the LB as shown in Fig.~\ref{fig:dpdk-concury}. On a single thread, \sysn can process and forward at least 17.63 Mpps (62.5\% of the maximum capacity). \textbf{We do not find a better single-thread software LB throughput in the literature.} Using 2 threads, \sysn can achieve the maximum network capacity of the node. As comparison, hash table based method cannot achieve the network capacity by 2 threads.
We expect much higher throughput of multi-thread \sysn if it is deployed on servers with more powerful NICs and memory buses. We tried some nodes in CloudLab with higher bandwidth NICs. They are either unavailable or has compatibility problems to DPDK. The results show that \sysn can achieve very high performance with little resource.

\begin{figure}[t]
    \centering
    \includegraphics[width=0.95\linewidth]{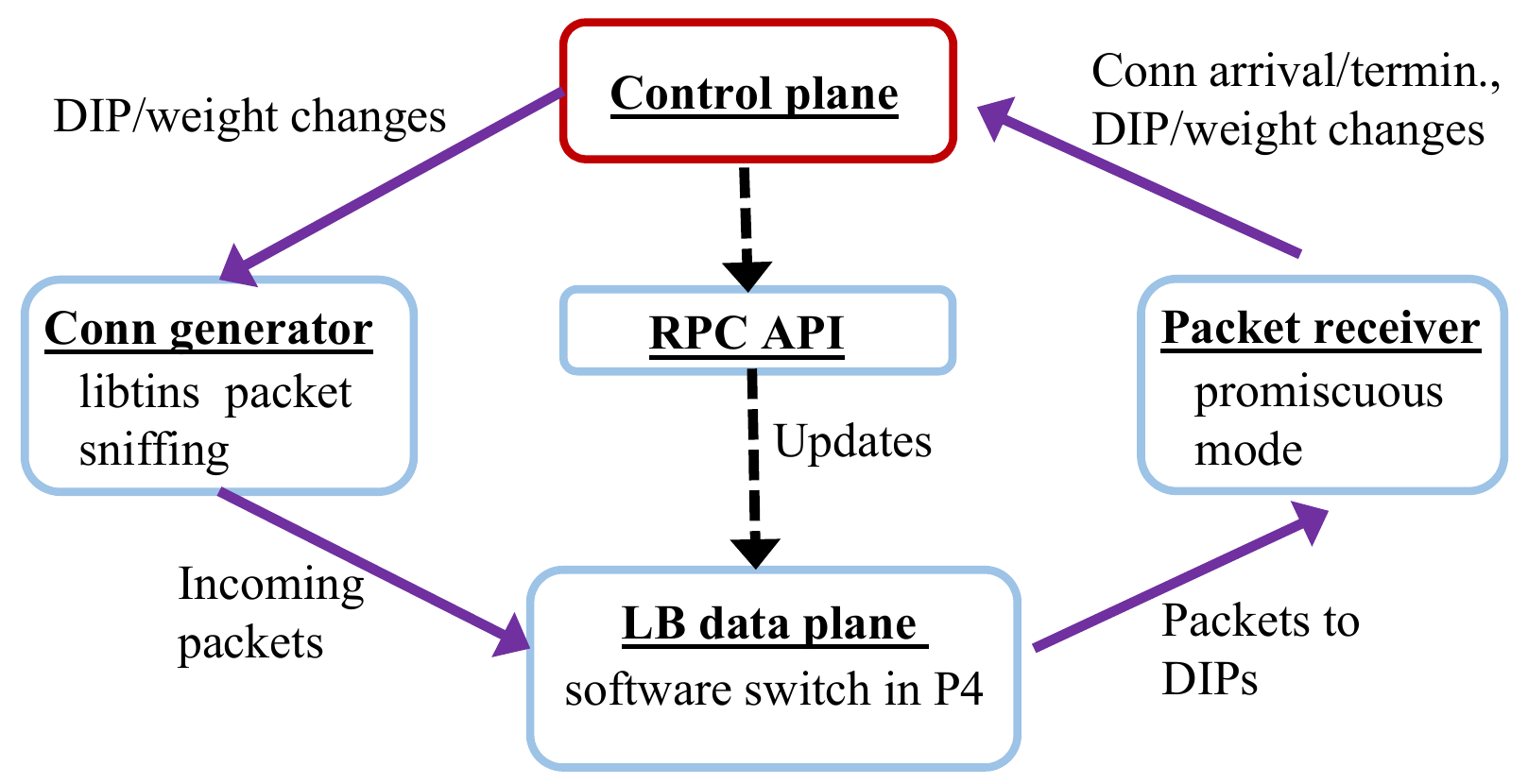}
    \caption{P4 prototype on Mininet}
    \label{fig:P4}
\end{figure}

\subsection{Evaluation on P4 prototype}
\label{sec:P4exp}

\textbf{P4 prototype of \sysn.}
We have also built a P4 prototype of \sysn, in which the  data plane includes around 400 lines of P4 code. The prototype is based on the simple switch behavioral model \cite{P4model} of the P$4_{16}$ language \cite{P416}. To manage the data plane tables, we add a middle layer between the data plane and control plane with C++ Thrift remote procedure call (RPC) API provided by the PI library \cite{PI}.

We use Mininet \cite{mininet} to implement the experimental platform to run \sysn, which includes a P4 switch as the \sysn LB, a \sysn control plane program, a host to generate packets from clients, and a host representing 16K logical DIPs, as shown in Fig.~\ref{fig:P4}. The receiving host uses the promiscuous mode to accept packets with different DIPs. We used libtins network packet sniffing library \cite{libtins} to generate and send packets. To allow the control plane to communicate with the data plane through RPC, we add the NAT support to the prototype hence the host can access TCP ports of the physical machine.

We use the P4 prototype to evaluate the load balancing of \sysn using both real and synthetic traffic.
Since the P4 prototype runs on Mininet, the throughput is of small importance because it does not reflect the actual throughput of a P4 switch. However, given that \sysn shows significant improvement over SilkRoad on software LB as shown in \S~\ref{sec:commodity} and \sysn-DP is no more complex than that of SilkRoad, we expect \sysn's throughput on a hardware switch may be no worse than that of SilkRoad.

In this set of experiments, every VIP has 128 DIPs and  DIPs have different weights, which reflect their resource capacities, to receive new connections. We use each connection to represent a state.  We define a metric $L$, called the normalized DIP load, as $L=c_i/w_i$ where $c_i$ is the number of connections forwarded to $DIP_i$ and $w_i$ is the weight of $DIP_i$. We show the normalized DIP load inside one VIP in Figures~\ref{fig:REALConnoverWeight} and \ref{fig:LFSRConnoverWeight} for real and synthetic traffic respectively. We use the same random seeds for experiments of both figures.  We find that the load for different DIPs are evenly distributed. Two DIPs showing 0 are with weight 0. The results of the other VIPs are very similar.

We show whether \sysn's load balancing will be affected by network dynamics in a new set of experiments. We allow new connections to come at a rate of 1K connections per VIP per second. In addition, every minute there is a big change in the weights of DIPs. We define the \emph{load balancing measure} as the maximum of all normalized DIP load over the average normalized DIP load: $\max(L_j)/ \mathrm{average}(L_j)$, where $L_j$ is the normalized load of DIP $j$. Hence, 1 means perfect load balancing. The results are shown in Fig.~\ref{fig:dynamic-load-balance-60}. We find that using \sysn, which is a weighted LB, the measure is always close to 1. When there are weight changes, the measure increases temporarily, because the new weights do not match the current connection distribution. However, it quickly returns back. If we use a non-weighted LB method,  such as SilkRoad \cite{SilkRoad}, the measure will be very high and result in bad load balancing. We further change the weights once very 30 second and show the results in Fig.~\ref{fig:dynamic-load-balance}. \sysn is still very resilient to weight changes.

We use Fig.~\ref{fig:concury-silkroad-balance-measure-cdf} to show the cumulative distribution of the DIP load at 2 seconds after the big weight change. We find all DIP loads of \sysn is between [0.75, 1.4], indicating good load balancing. The DIP loads will quickly converge to 1 from Fig.~\ref{fig:dynamic-load-balance}. Some servers of Silkroad may experience high load ($>4$), which may cause server overloading.

\subsection{Summary of evaluation}
As stated in \S~\ref{sec:problem}, our design objectives include high packet processing throughput, efficiency of memory cost, weighted load balancing, quick construction, and packet consistency.  \sysn performs well in all aspects. Compared to prior solutions, \sysn shows the advantages in all these aspects and is only weaker in inserting new states as shown in Fig.~\ref{fig:vary-dip-count-static-add}, which is still sufficiently good for large cloud networks. In addition, \sysn is a portable solution and does not rely on any specific platform.

\section{Conclusion}
\label{sec:conclusion}
We design and implement a new software stateful LB called \sysn, which achieves weighted balancing of incoming traffic, maintaining consistency, high throughput, memory efficiency, and false hit freedom. It satisfies the requirements of the load balancer for cloud and  edge  data centers. \sysn makes use of the theoretical studies of minimal perfect hashing and apply a compact data structure that represents the concurrent states without storing the actual state information. We implement \sysn on both software and P4 prototypes. Evaluation results show that  \sysn provides higher packet processing throughput by $>$2x and lower memory cost compared to existing stateful LB algorithms. Our future work will be extending \sysn to mobile client environments.

	\small
	\bibliographystyle{abbrv}
	\bibliography{paper}
	\appendix
	\section{Appendix}
\subsection{A. Example of consistency violation by static hashing.}
Consider  the example shown as Fig.~\ref{fig:DIPchange}. Suppose the LB uses static hashing to evenly distribute traffic to four DIPs.  Connection $C_1$, whose hash value is 0.3, is mapped to $DIP_2$. All packets of $C_1$ should be forwarded to $DIP_2$ if there is no DIP pool change. However, if there is a change of the DIP pool, e.g., the failure of $DIP_4$, then the hashing-to-DIP mapping need to be adjusted for balancing. As a result, later packets of $C_1$ will be forwarded to $DIP_1$, causing a PCC violation. Other stateless hashing algorithms such as consistent hashing experience similar problems. These problems are more significant in edge networks where the state may be multi-connection and long-term.

\begin{figure}[h]
    \centering
    \includegraphics[width=0.99\linewidth]{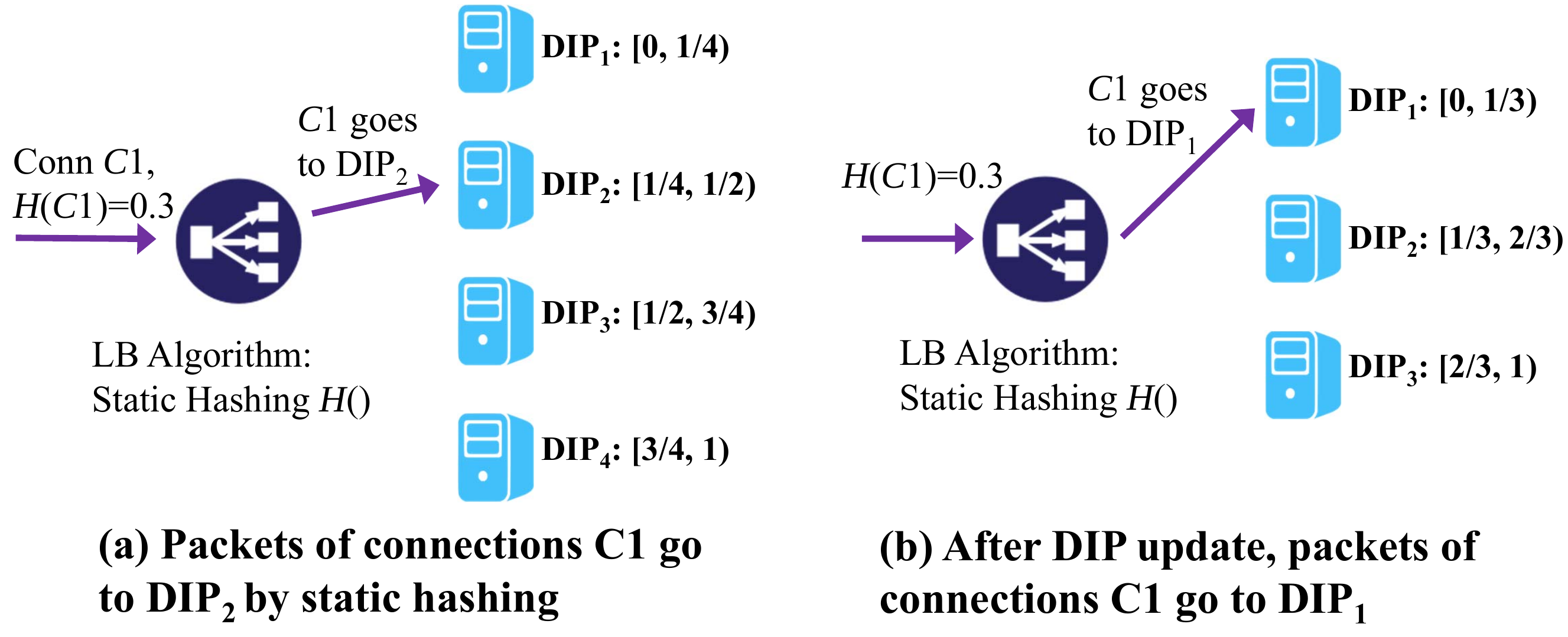}
    \caption{PCC violation of static hashing}
    \label{fig:DIPchange}
\end{figure}

\subsection{B. Pseudocode}

We also show the pseudocode of the \sysn-DP lookup algorithm in  Algorithm~\ref{alg:lookup} and the \sysn-DP updating algorithm in Algorithm~\ref{alg:update}.

{\small
\begin{algorithm}[t]
 \SetKwInOut{Input}{Input}\SetKwInOut{Output}{Output}
 \Input{VIP index $i$, 5-tuple $t$, hash func. $h_a$ and $h_b$}
 \Output{DIP $d$}
 \nl $A_i \gets VIPArray[i]$\;
 \tcp{$A_i$:\,memory address of array\,$A$ of the $i$-th\,Othello}
 \nl $B_i \gets A_i+m_a$\;
 \tcp{$B_i$:\,memory address of array\,$B$ of the $i$-th\,Othello}
 \nl $Dcode \gets A_i[h_a(t)]\oplus B_i[h_b(t)]$\;
\nl $d \gets DA[i][Dcode]$\;
 \caption{Data plane lookup algorithm of \sysn}
 \label{alg:lookup}
\end{algorithm}
}

{\small
\begin{algorithm}[t]
 \SetKwInOut{Input}{Input}\SetKwInOut{Output}{Output}
 \Input{$\langle i, A', B', DA' \rangle$ from update message}
\tcp{$i$: VIP index; $A'$: new array $A$; $B'$: new array $B$, $DA'$: new $DA$ in dimension $i$}
 \nl $A_i \gets VIPArray[i]$\;
 \nl $B_i \gets A_i+m_a$\;
\nl $ArrayCopy(A_i, A')$; $ArrayCopy(B_i, B')$\;
\nl $ArrayCopy(DA[i][], DA')$\;
\tcp{$ArrayCopy$ copies received arrays to existing ones with concurrent control.}
 \caption{Data plane update algorithm of \sysn}
 \label{alg:update}
\end{algorithm}
}

\subsection{C. Data plane complexity analysis and comparison.}
\textbf{Time cost.} \emph{1) \sysn.} \sysn-DP is very simple and fast. Each lookup is in $O(1)$, including \textit{at most} 6 read operations from static arrays, 2 hash computations (32 bits for each), and an XOR computation. The 6 read operations include 1 for the VIP array access, 1 for basic information of the Othello, 2 for Othello access, and 2 for finding the actual DIP of the $Dcode$.
The time cost is the same for stateful and stateless packets.
\emph{2)Cuckoo$+$digest.}
We compare \sysn with the  hash table plus digest approach, which is applied by some main stream systems \cite{Maglev,SilkRoad}. We assume a (2,4) Cuckoo hash table which has been shown as an optimized and up-to-date LB design choice of a hash table \cite{SilkRoad}.
For a stateful packet, \emph{in average} Cuckoo$+$digest needs 3.5 hash computations, including 2 for generating the 64-bit digest and 1.5 for locating the buckets (50\% found in the first bucket and 50\% found in the second bucket). It also takes 7 memory read operations in average: 1 for basic information of the hash table, and 6 for hash table lookups (4 lookups per bucket). It takes 6 digest comparisons in average (4 per bucket).
For a stateless packet, Cuckoo$+$digest needs to read and compare the key digest to all slots in the two buckets, hence it takes 4 hash computations, 9 memory read operations, and 8 digest comparisons.
\sysn is faster compared to Cuckoo hashing based solutions as shown in Table~\ref{tbl:dp-time}.

\newcolumntype{C}[1]{>{\centering\let\newline\\\arraybackslash\hspace{0pt}}m{#1}}
\begin{table}
 \centering \small
  \begin{tabular}{C{2.8cm}C{1.2cm}C{1cm}C{1.5cm}}
     \toprule
  LB Algorithm   &\#Hashes&\#Reads&Other computation\\
     \hline
Cuckoo$+$digest \cite{SilkRoad,Maglev} stateless pkts & 4 & 9  & cmpr digest 8 times \\
     \hline
Cuckoo$+$digest \cite{SilkRoad,Maglev} stateful pkts & 3.5 & 7  & cmpr digest 6 times \\
     \hline
\sysn & 2 & 6 & 1 XOR  \\
\bottomrule
\end{tabular}
  \small\caption{Data plane time cost breakdown (per lookup)}
\label{tbl:dp-time}
\end{table}

\textbf{Space cost.} \emph{1)\sysn.} Let $n$ be the number of total states and $l_d$ is the length of $Dcode$, assuming all Othellos use the same length of $Dcode$. The Othellos take $2.33l_d*n$ bits, the VIP array takes $64m$ bits, and the DIP array takes $2^{l_d}l_vm$ bits where $l_v$ is the length of the DIP index. A DIP and port take 48 bits. The total space cost is $2.33l_dn+64m+2^{l_d}l_vm+48*2^{l_v}$ bits.
\emph{2)Cuckoo$+$digest.}
Assume the hash table load factor is 90\%, Cuckoo$+$digest takes $1.1*(64+l_v)*4n$ for the hash table, $2^{l_d}l_vm$ for the weighted load balancer \cite{Maglev}, and $48*2^{l_v}$ for the DIP retrieval table. Hence the total is
$1.1*(64+l_v)n+2^{l_d}l_vm+48*2^{l_v}$ bits.
Since $n\gg m$, to compare the space cost of \sysn and Cuckoo$+$digest is mainly comparing  $2.33l_dn$ and $1.1*(64+l_v)n$. In practical settings, $2.33l_dn$ is much smaller than $1.1*(64+l_v)n$. For example, using $l_d=12$ and $l_v=12$, $2.33l_dn=28n$ and  $1.1*(64+l_v)n=83.6n$. The experimental results show that the Cuckoo$+$digest method (Maglev) needs around 3x memory compared to \sysn, which agrees with the analysis here. Note one assumption here is that all VIPs use a same length of  $Dcode$.

\subsection{D. Example of OthelloMap.}

\begin{figure}
    \centering
    \includegraphics[width=0.99\linewidth]{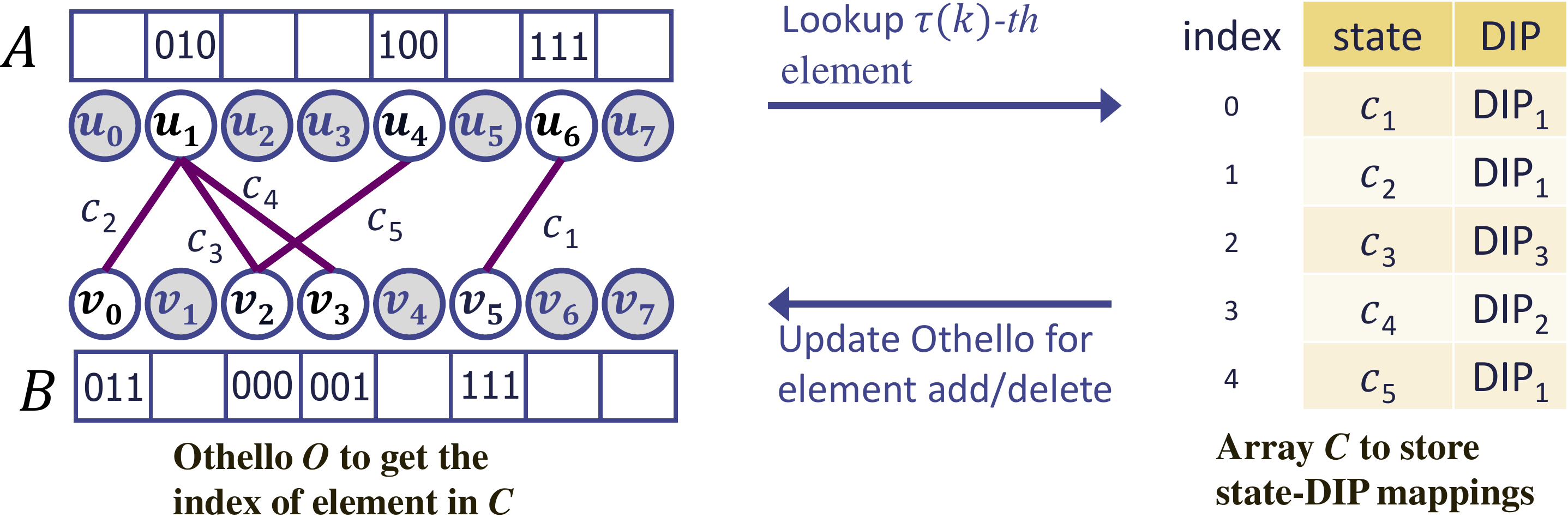}
    \caption{OthelloMap of 5 state-DIP mappings}
    \label{fig:OMap}
\end{figure}

Figure~\ref{fig:OMap} shows an example of OthelloMap for one VIP. The array $C$ stores all current state-DIP mappings. OthelloMap also maintains an Othello structure to reflect the index of each mapping in the array. For example, state $c_1$ is stored at index 0 of the array. Hence the lookup result of $c_1$ in the Othello $O$ is $111\bigoplus 111=0$. Once a new state-DIP mapping is inserted or an expired mapping is deleted from $C$, $O$ should change accordingly. If an existing mapping, say $c_2$ to $DIP_1$ at index 1 is deleted, the mapping as the last element, i.e., $c_5$ to $DIP_1$ should be moved to index 1.


\end{document}